\documentclass[aps,twocolumn, superscriptaddress, nofootinbib]{revtex4-1}

\usepackage[usenames]{color}
\usepackage{xcolor}
\usepackage{upgreek}
\usepackage{amsmath}
\usepackage{amssymb} 
\usepackage{graphicx}
\usepackage[utf8]{inputenc}
\usepackage{physics}
\usepackage{empheq}
\usepackage{braket}
\usepackage{mathrsfs}
\usepackage{siunitx}
\usepackage{float}
\usepackage{tcolorbox}
\usepackage{enumitem}
\usepackage{placeins}
\usepackage{xcolor}
\usepackage[T1]{fontenc}
\DeclareMathSizes{10}{10}{10}{10}

\begin{document}
\title{Embodying computation in nonlinear perturbative metamaterials} 
\author{Sima Zahedi Fard}
\affiliation{AMOLF, Science Park 104, 1098 XG Amsterdam, the Netherlands}
\author{Paolo Tiso}
\affiliation{Institute for Mechanical Systems,  ETH Zürich, 8092 Zürich, Switzerland}
\author{Parisa Omidvar}
\affiliation{AMOLF, Science Park 104, 1098 XG Amsterdam, the Netherlands}
\author{Marc Serra-Garcia}
\affiliation{AMOLF, Science Park 104, 1098 XG Amsterdam, the Netherlands}

\date{\today}

\begin{abstract}

\end{abstract}

\maketitle

\textbf{Designing metamaterials that carry out advanced computations poses a significant challenge. A powerful design strategy splits the problem into two steps: First, encoding the desired functionality in a discrete or tight-binding model, and second, identifying a metamaterial geometry that conforms to the model. Applying this approach to information-processing tasks requires accurately mapping nonlinearity---an essential element for computation---from discrete models to geometries. Here we formulate this mapping through a nonlinear coordinate transformation that accurately connects tight-binding degrees of freedom to metamaterial excitations in the nonlinear regime. This transformation allows us to design information-processing metamaterials across the broad range of computations that can be expressed as tight-binding models, a capability we showcase with three examples based on three different computing paradigms: a coherent Ising machine that approximates combinatorial optimization problems through energy minimization, a mechanical racetrack memory exemplifying in-memory computing, and a speech classification metamaterial based on analog neuromorphic computing. }

\begin{figure}[b]
    \centering
    \includegraphics[width=\linewidth]{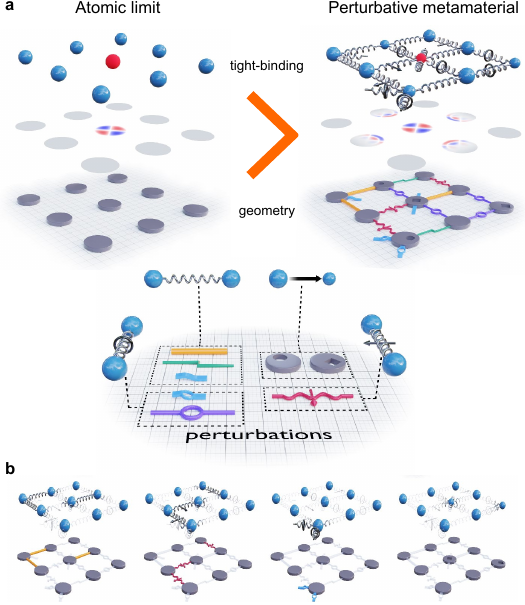}
    \caption{\textbf{| Embodying a nonlinear tight-binding model in a perturbative metamaterial}  \textbf{a} We identify tight-binding degrees of freedom (red sphere) with modes of the individual resonators and construct the target model by `mixing-and-matching' geometric perturbations, which realize linear springs, various types of nonlinear interactions, and change the site frequency. \textbf{b} Perturbations have a local effect on the tight-binding model.}
\label{fig:concept}
\end{figure}

The incorporation of nonlinearity in elastic metamaterials has unlocked a host of novel information-processing capabilities---from counting and recognizing sequential patterns~\cite{PhysRevLett.130.268204}, to implementing the control algorithm of a robot in the elastic domain~\cite{rafsanjani2019programming}. Performing these computations with metamaterials provides inherent advantages of parallel operation~\cite{mousa2024parallel}, biocompatibility~\cite{tang2024injectable}, low power consumption~\cite{luo2025wireless}, and direct processing of signals in their native domain without electronic transduction~\cite{tenaspeech}.  However, tackling general, complex computation tasks in a metamaterial poses significant, unique challenges: Information-processing tasks prescribe distinct responses to distinct inputs, causing the space of possible behaviors to grow exponentially with input size---making the direct optimization~\cite{bordiga2024automated} of metamaterial responses unfeasible. This challenge is compounded by the fact that specifying the desired input-output map a priori often entails solving the very computational problem one seeks to embody in the material.




While designing metamaterial geometries is hard, coming up with discrete (mass-spring or tight-binding) models that compute is much easier. Documented examples include models that perform speech recognition~\cite{coulombe2017computing, bohte2025mass}, solve combinatorial optimization problems~\cite{heugel2022ising}, sample diffusion models~\cite{bosch2025local}, train neural networks~\cite{de2025learning}, or even realize an entire 8-bit CPU~\cite{serra2019turing}. In this work, we introduce a scalable method to design geometries from nonlinear tight-binding models, allowing us to embody complex tasks---traditionally the exclusive domain of digital computers---into elastic metamaterials. 

\section*{Results}
Our design approach is based on perturbative metamaterials~\cite{Serra-Garcia2018, Matlack2018, tenaspeech}---consisting of weakly interacting sites. These metamaterials can be seen as a perturbation on an atomic limit of disconnected resonators, allowing the identification of tight-binding degrees of freedom with local modes of resonator sites (Fig.~\ref{fig:concept}a). Perturbative metamaterials are highly effective at embodying tight-binding models because the effect of geometric perturbations is approximately local and additive. Locality means that interactions affecting each tight-binding degree of freedom depend only on geometric features in the vicinity of the corresponding site (Fig.~\ref{fig:concept}b). Additivity allows us to reason about geometric elements as having an independent effect on the tight-binding model, drastically simplifying the exponentially large space of combinations of geometric features. These properties combine to form a benign design problem: First, the tight-binding degrees of freedom are mapped to local orbitals of the metamaterial sites. Then, the tight-binding model is approximately embodied by `mixing-and-matching' a small set of geometric perturbations, each of them corresponding to a few tight-binding potential terms. Finally, the geometry can be refined using gradient descent to accurately match the target model~\cite{Matlack2018}. This approach has been remarkably successful, enabling the realization of higher-order topological insulators~\cite{Serra-Garcia2018}, non-Hermitian phases~\cite{FAN2022108774} and convolutional neural layers~\cite{tenaspeech}.

\begin{figure}[t]
    \centering
    \includegraphics[width=\linewidth]{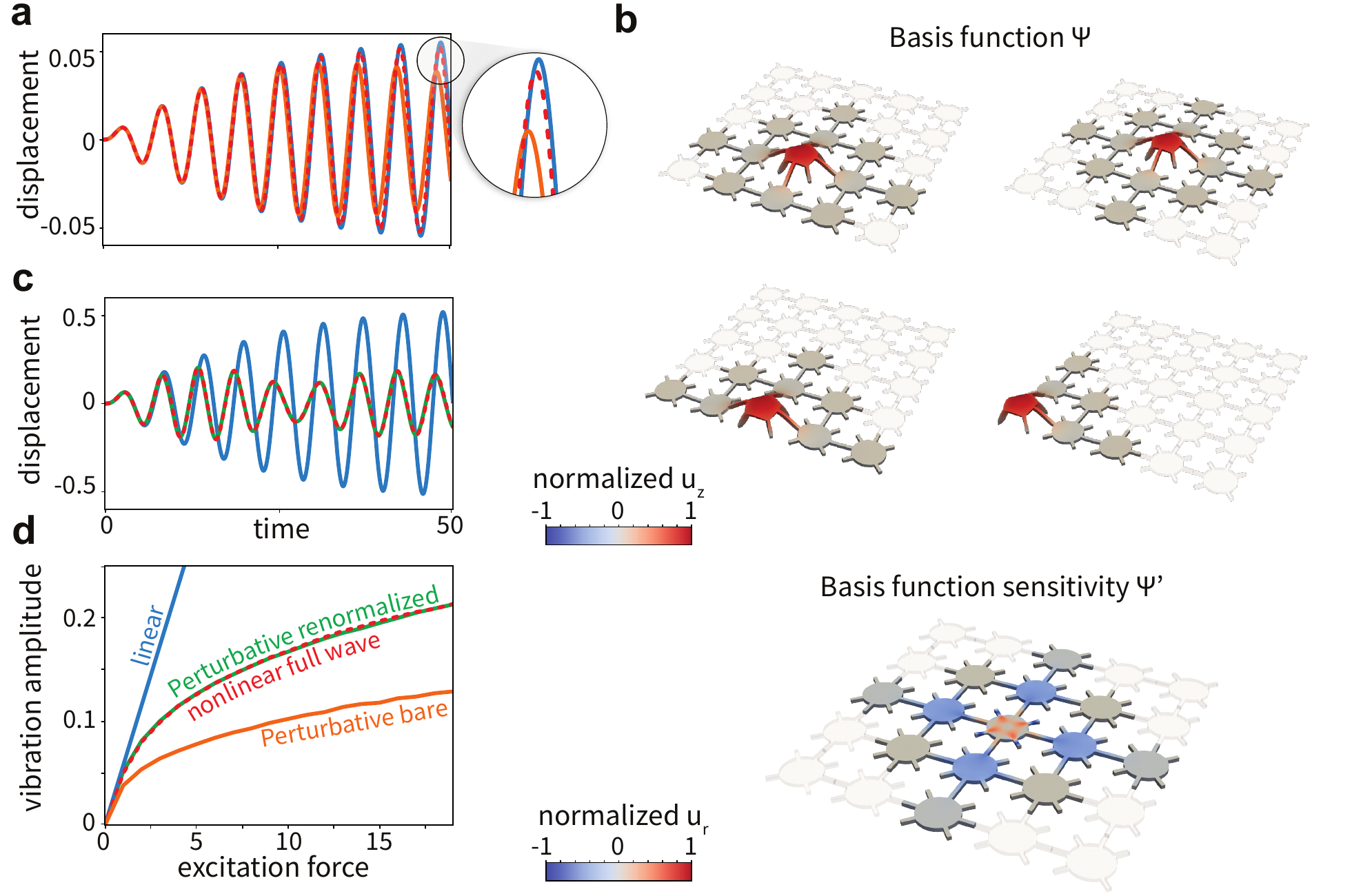}
    \caption{\textbf{| Perturbative metamaterials in the nonlinear regime}  \textbf{a} In the small-amplitude transient response, a linear approximation (blue) is closer to the full-wave simulation (dashed red) than a bare nonlinear perturbative model without the coordinate transformation of Eq.~\ref{eq:NonlinearCoordTransformation} (red) \textbf{b} Basis functions and basis function sensitivities. In the basis function, the color represents the out-of-plane displacement $u_z$, while in the basis function sensitivity, the color represents the in-plane radial displacement $u_r$. The basis functions are computed in small clusters (dark sites), consequently the computation can be performed in parallel and at a constant cost per site. \textbf{c} Transient response computed with full wave simulation (dashed red), linear perturbative model (blue), and nonlinear perturbative model with coordinate transformation (green). With the nonlinear coordinate transformation, the perturbative model accurately captures the nonlinear response. \textbf{d} Steady-state amplitude computed via full wave simulation (dashed red) and via perturbative linear (blue), bare nonlinear (orange), and coordinate-transformed (green) models. }
\label{fig:NonlinearOscillation}
\end{figure}

Although highly successful for linear problems, perturbative metamaterials do not naturally extend to the nonlinear regime. In some cases, naively incorporating nonlinearity can result in less accurate predictions than ignoring nonlinearity altogether (Fig.~\ref{fig:NonlinearOscillation}a). This failure originates in the map between tight-binding and metamaterial degrees of freedom. In perturbative metamaterials, tight-binding degrees of freedom are mapped to localized excitations with a fixed shape. However, in the presence of nonlinearity, the shape of these excitations becomes deformation-dependent. We address this by using a nonlinear coordinate transformation that captures the basis function changes introduced by the nonlinearity~\cite{JAIN201780, rutzmoser2017generalization, idelsohn1985load}, ensuring that the tight-binding model maps to the correct effective theory around every instantaneous deformed configuration,
\begin{equation}
u(x,t)=q_i(t)\Psi_i(x)+\frac{1}{2}q_i(t)q_j(t)\Psi_{ij}'(x),
\label{eq:NonlinearCoordTransformation}
\end{equation}
where $u$ is the metamaterial deformation field, $q_i$ are tight-binding degrees of freedom, $\Psi_i$ are the localized basis functions and $\Psi'_{ij}$ quantifies the sensitivity of the $i$-th basis functions to deformations of the $j$-th degree of freedom (Fig.~\ref{fig:NonlinearOscillation}b). The coordinate transformation captures the deformation of all higher-order modes in response to nonlinear forces (see Supplementary Information), and is accurate, to second order, as long as there is a time scale separation between basis functions and higher-order modes. This requirement, although a limitation, can be satisfied by making use of the design freedom to pick up a geometry with the required time scale separation~\cite{suri1989design}. 

Equipped with a map between geometry and nonlinear tight-binding model that is accurate (Fig.~\ref{fig:NonlinearOscillation}c, d) and presents a well-posed design problem, we will exemplify our approach through the embodiment of three nonlinear information-processing metamaterials.

\subsection*{Example 1: Metamaterial Coherent Ising Machine}
%
\begin{figure}[b!]
	\centering
	\includegraphics[width=1.0\linewidth]{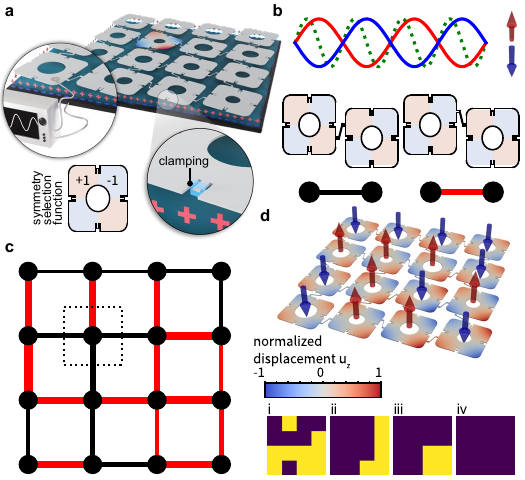}
	\caption{\textbf{| Metamaterial coherent Ising machine} \textbf{a} Metamaterial design. Every side of the unit cell is clamped at the mid-point, so that the first resonance mode presents both positive and negative boundary displacements---allowing for the realization of positive and negative interaction terms. The metamaterial is subject to a time-varying electric field at $2\omega_0$, where $\omega_0$ is the plate resonance frequency.  \textbf{b} The stiffness modulation (green, dotted) results in two possible stable phase oscillations (red, blue), that we call \emph{spin-up} and \emph{spin-down}. Depending on the location of the connecting bars, the plates interact via ferromagnetic (black line) or antiferromagnetic (red line) couplings. \textbf{c} Tight-binding model of the system.  \textbf{d} Metamaterial displacement at the end of the simulation. The arrows indicate the phase of oscillation, which corresponds to a ground state of the Ising model in panel c. The subplots show the product between the phases of the system and a ground state of the model in panel c at times $t=16T$, $t=60T$, $t=80T$ and $t=160T$, where $T$ is the period of oscillation.} 
	\label{fig:cim}
\end{figure}
Coherent Ising Machines (CIMs)~\cite{marandi2014network, casilli2024parametric, mcmahon2016fully, cilasun2025coupled} are physical solvers for combinatorial optimization problems. They are rooted on the observation that the ground state of an Ising Hamiltonian corresponds to the solution of a Quadratic Unconstrained Binary Optimization (QUBO) problem; since the QUBO problem is NP-hard, a physical system that determines the (approximate) ground state of an Ising Hamiltonian can be used to find (approximate) solutions to any problem in NP. CIMs encode an Ising Hamiltonian in a network of degenerate parametric oscillators (Fig.~\ref{fig:cim}a), that support two stable phases of oscillation and can thus be assimilated to a spin system (Fig.~\ref{fig:cim}b). 
\\
\\
We construct a CIM described by the tight-binding model with potential per site $i$~\cite{heugel2022ising} given by
\begin{equation*}
V_i=\frac{1}{2}\left( \omega_0^2+\alpha \cos(2\omega_0 t) \right)q_{i}^2  + \frac{1}{4}\lambda q^4_{i} -\sum_{j\in nn(i)}{J_{ij}q_iq_j},
\label{eq:cim}
\end{equation*}
where q are tight-binding degrees of freedom, $nn(i)$ are the nearest-neighbors, $\omega_0$ is the natural frequency of the site, $\alpha$ is the parametric drive, $\lambda$ is a Kerr (Duffing) nonlinearity---responsible from preventing the amplitude to grow exponentially, and $J_{ij}$ is the coupling matrix that stores the specific Ising problem, here set to encode a frustration-free problem.

We map the tight-binding model to a metasurface consisting of square plates, held by clamped bars at the middle of each side (Fig.~\ref{fig:cim}a). This causes the first lattice mode to contain both positive and negative boundary displacements, and thus allows for positive and negative values of $J_{ij}$. The parametric drive is introduced by a time-dependent electric field---modelled as a time-dependent surface stiffness. Once the model has been approximately embodied by setting the couplings, we perform gradient descent on the plate holes to achieve the correct resonance frequency. The final effective model, extracted from the geometry, is shown in Fig.~\ref{fig:cim}c. We validate the design by simulating the structure using a nonlinear full wave simulation (see Methods), and observe that it converges to a ground state of the system (Fig.~\ref{fig:cim}d). Since the effective Ising model does not include a local magnetic field, the system has two degenerate ground states $\vec{\sigma}_0$ and $-\vec{\sigma}_0$. Any state of any spin can be seen as belonging to one of the ground states.  Initially, the system does not show preference for either ground state (Fig.~\ref{fig:cim}d). As the simulation evolves, one of the domains grows until it encompasses the entire device, at which point a solution to the QUBO problem has been obtained.

\begin{figure}[h!]
	\centering
	\includegraphics[width=1.0\linewidth]{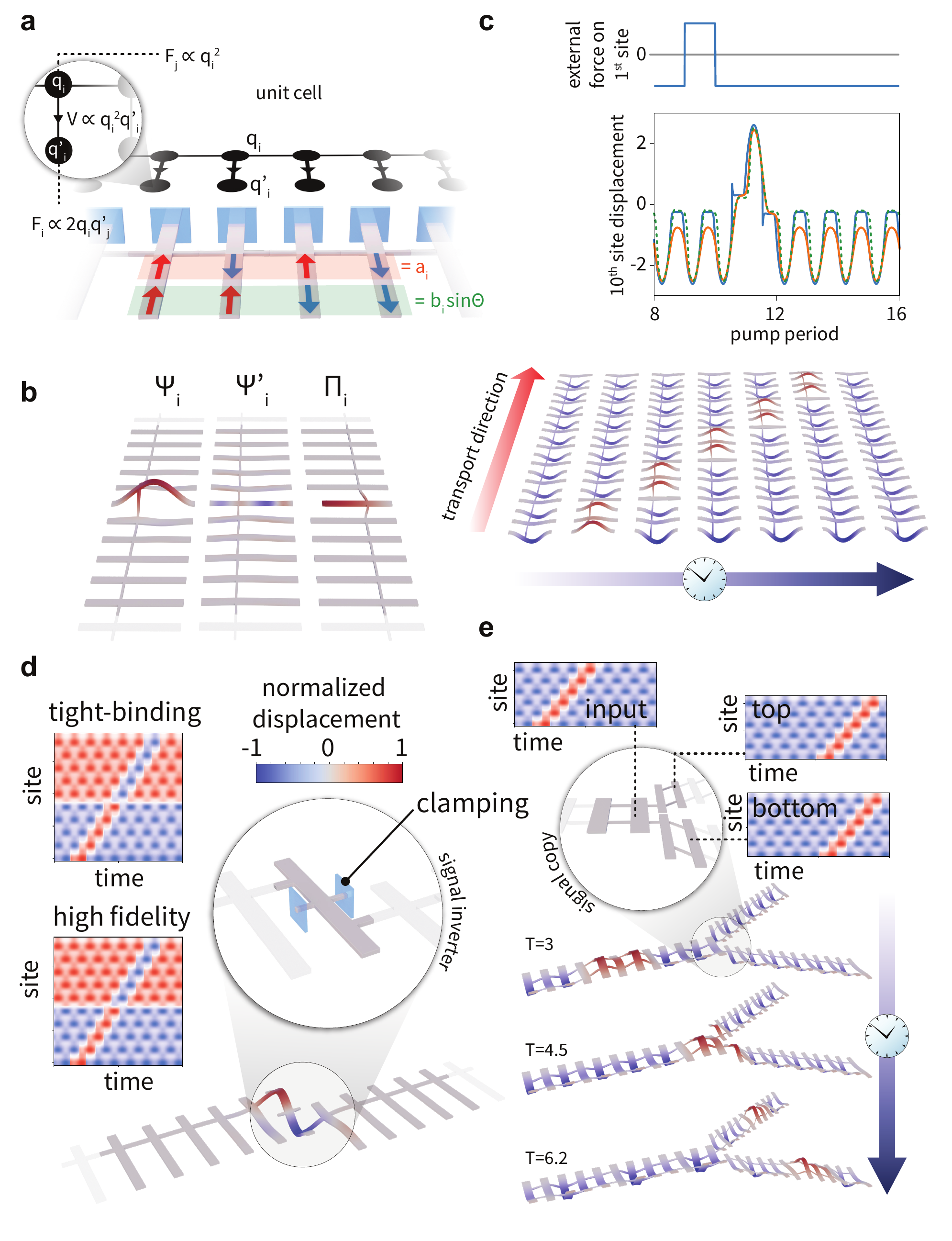} 
    \caption{Quantized transport of topological solitons | \textbf{a} Nonlinear tight-binding model (top) and device geometry (bottom). Each site $i$ is modelled by two degrees of freedom, representing buckling ($q_i$) and boundary ($q_i'$) deformations respectively. \textbf{b} Localized basis functions corresponding to buckling deformation ($\Psi$), boundary compression ($\Pi$), and the corresponding self-modal derivative associated to the buckling eigenmode ($\Psi'$). \textbf{c} Top: A pulsed force on the first site produces a topological soliton, that propagates when the boundary compressions are modulated. The trajectory of the tenth beam is calculated via overdamped simulation of the original tight-binding model (solid, orange), via a high fidelity nonlinear equilibrium continuation (blue), and via tight-binding model incorporating long-range interactions (green dotted line). Bottom: 3D representation of the soliton propagation. \textbf{d} Top-left: Lattice incorporating an inverting element. The central supporting arm is fixed. Displacements of the beam centers as a function of time, for an inverting chain, computed using the tight-binding model including long-range interactions and using the high-fidelity nonlinear equilibrium continuation on the full finite element model. In the bottom,  localized basis function associated to a buckling deformation of the inverting beam. The basis function alternates positive and negative signs, allowing for signal inversion.  \textbf{e} Branched geometry, implementing a signal copy, simulated via tight-binding model including long-range interactions (color plots) and full nonlinear equilibrium continuation (3D images)} 
    \label{fig:topologicalSoliton}
\end{figure}

\subsection*{Example 2: Elastic racetrack memory}
In traditional computer architectures, memory and processing are separate---resulting in a large energy cost known as the von Neumann bottleneck. Racetrack memories~\cite{hayashi2008current, parkin2008magnetic, pham2024fast, heydermanDomainWallLogic} are a form of in-memory computing where persistently-stored information (for example, in magnetic domains) is transported along a circuit by the action of external currents or fields. Although racetrack memories have been largely the domain of magnetic phenomena, we have found that similar operation can be achieved in generic multistable systems using topological edge mode instabilities to drive the transport of information (see [ref] for an experimental realization, discussion on the physical mechanism, and application to in-memory logic circuits). The mechanical racetrack memory is described by the tight-binding energy
\begin{equation*}
V_i=\frac{1}{2}{\omega_0^2}q^2+\gamma q_i'q^2+\frac{1}{4}\lambda q_i^4-c{q_iq_{i+1}}, 
\label{eq:TopoRatchet}
\end{equation*}
where $q$ represents the field variable, $q'$ is an externally-controlled boundary compression field, $\omega_0^2$ is the local potential, $c$ is the nearest-neighbor hopping strength, $\gamma$ quantifies the second-order nonlinear interaction between the field and the boundary conditions and $\lambda$ is the Kerr nonlinearity. A schematic of the effective theory is shown in Fig.~\ref{fig:topologicalSoliton}a. The geometry is periodic on a single site, but a lattice with an effective unit cell of four sites will be constructed by prescribing the external compression field $q'$ following the pattern $q_i'=a_i+b_i\sin (\theta)$ with $a_i=(1,-1,1,-1)$ and $b_i=(1,1,-1,-1)$. In this model, sites with a negative local potential ($a_i<0$) are bi-stable and persistently store information, which is transported by a unit cell every time $\theta$ is increased by $2\pi$.

We embody the tight-binding model on a lattice of interacting beams (Fig.~\ref{fig:topologicalSoliton}a), connected through thin arms. We map the field $q_i$ to the first vibration mode of the beam $i$, $\Psi_i$---which will become the first buckling mode once a compression is applied, providing the bi-stability necessary for persistent information storage---incorporating one modal sensitivity $\Psi_{i}'$ in the nonlinear coordinate transformation (Eq.~\ref{eq:NonlinearCoordTransformation}). The function $\Psi_{i}'$ captures the distortion of the mode $\Psi_i$ in response to a displacement $q_i$. The field $q_i'$ is mapped to a function $\Pi$ quantifying the deformation of the material to a unit boundary compression applied at each beam $i$ (Fig.~\ref{fig:topologicalSoliton}b). The resulting structure displays the expected quantized information transport (Fig.~\ref{fig:topologicalSoliton}c), although we observe a disagreement in the beam trajectories around their minimum displacement positions (Fig.~\ref{fig:topologicalSoliton}c). This deviation originates in long-range interactions that emerge due to imperfect localizations of the basis functions, a phenomenon that has also been observed in linear perturbative metamaterials~\cite{Serra-Garcia2018} and can be mitigated through unit cell geometry optimization~\cite{tenaspeech}. Incorporating these long-range interactions in the tight-binding model accurately reproduces the trajectory. Although not used here, long-range interactions could also be leveraged to implement advanced tight-binding models.

We can construct circuit elements with mechanical racetracks. Digital inverters---an essential ingredient for digital logic---can be designed by incorporating negative couplings, engineered by clamping the beam at the midpoint (Fig.~\ref{fig:topologicalSoliton}d). This causes the first mode of buckling to present regions of positive and negative out of plane displacement. The negative coupling is realized by connecting a beam to the negative displacement region. Branched models can also be embodied (Fig.~\ref{fig:topologicalSoliton}e), providing the building blocks to realize complex logical circuits. Although perturbative metamaterials were originally introduced to model linear, narrowband vibrations, these results show that they can also be applied to quasistatic nonlinear multistable systems. Signals through the racetrack do not experience attenuation, as they get energy from the external drive. Thus, such system can be used to realize long digital circuits in elastic metamaterials---overcoming multiple challenges of the field, including attenuating signals~\cite{song2019additively} and the need to re-set the system after every computation~\cite{PhysRevLett.130.268204, mei2021mechanical}.

\subsection*{Example 3: Speech classification via reservoir computing}
\label{section:speech}
 \begin{figure}[b!]
	\centering
	\includegraphics[width=1.0\linewidth]{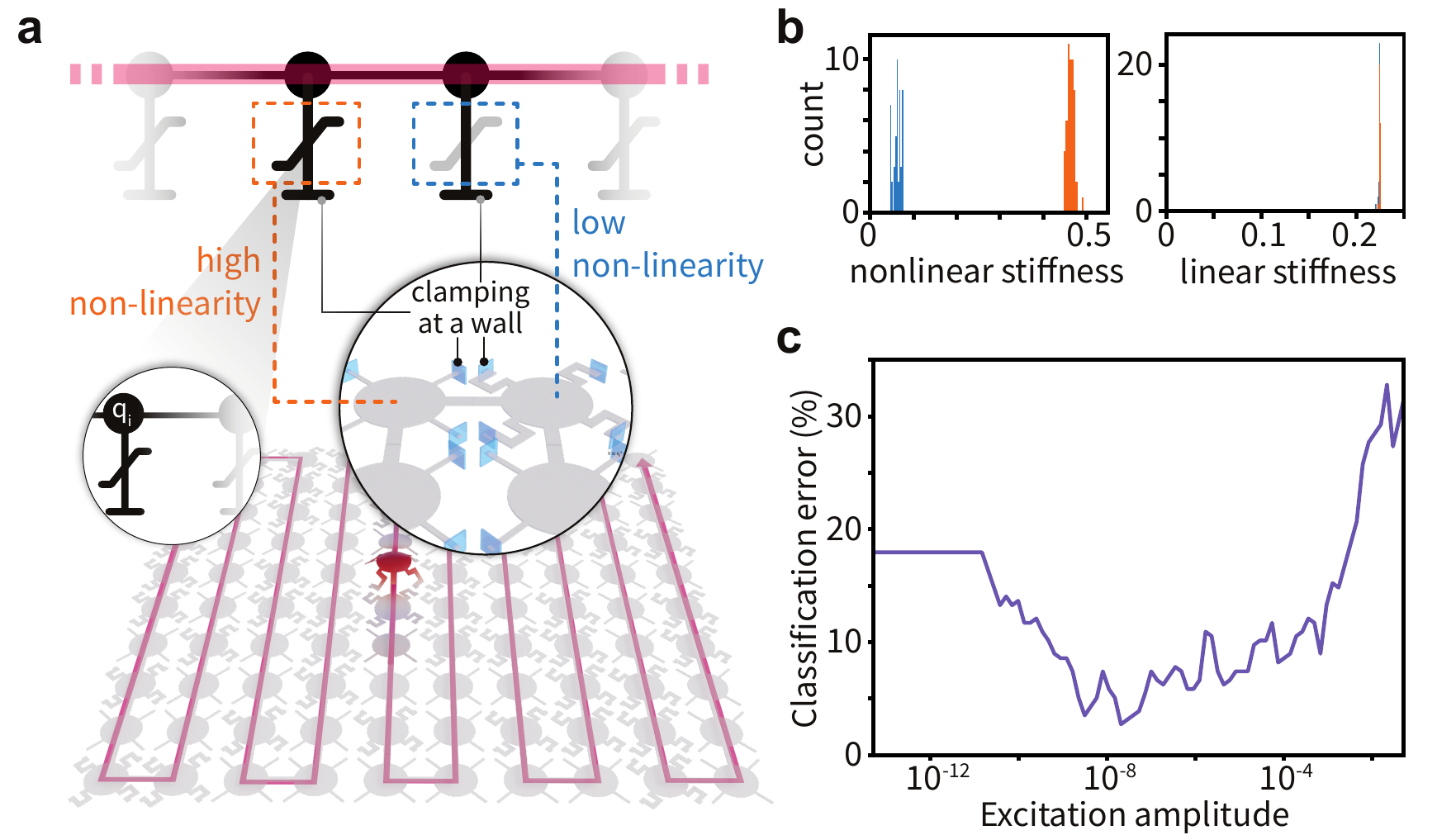}
	\caption{ \textbf{ | Reservoir computing with a nonlinear perturbative metamaterial} \textbf{a} Tight-binding model of the computing metamaterial. The material consists of a network of sites subject to a local potential, which can take two different values (low and high). These are realized in the drum network using two different support beam geometries. \textbf{b} Histogram of the local nonlinear term $\lambda$ (left) and linear potential term $\omega^2$ (right) for the C-shaped arms (blue) and straight arms (orange). The design degree of freedom allows us to realize different local nonlinearities while keeping the linear term constant. \textbf{c} Classification error as a function of the excitation amplitude. The metamaterial can perform speech classification. For low amplitudes it operates on the linear regime and the accuracy does not depend on amplitude, however when the amplitude exceeds the nonlinear potential the precision drastically increases. If the amplitude is increased beyond a limit value, the response becomes chaotic and the precision degrades again. }
 \label{fig:speech}
\end{figure}
A promising application for metamaterial computers is the processing of wave signals in their natural domain, for example to realize optical image enhancement or mechanical speech recognition. Metamaterials eliminate the need for transducers~\cite{tenaspeech} and can provide unprecedented advantages in speed and efficiency~\cite{luo2025wireless}. This realization has led to the proposal of several metamaterial designs for speech classification~\cite{doi:10.1126/sciadv.aay6946, tenaspeech}. So far, however, these designs have been limited to the linear regime, and therefore suffer from limited accuracy. Nonlinear speech classification has been demonstrated only in discrete models~\cite{coulombe2017computing, bohte2025mass}, but not realized as a metamaterial geometry. 

In this section, we will embody the effective model of Ref.~\cite{coulombe2017computing}---a nonlinear speech classifier---in a nonlinear metamaterial geometry. The model is based on reservoir computing~\cite{jaeger2001echo, maass2002real}, leveraging the complex response of a multi-DOF nonlinear system though an engineered readout function that combines the measured trajectories to maximize the classification accuracy (see methods). The model is governed by the potential energy per site $i$
\begin{equation*}
V_i=\frac{1}{2}{\omega_0^2}q^2_i+\frac{1}{4}\lambda q_i^4-c{q_iq_{i+1}}. 
\label{eq:SpeechModel}
\end{equation*}
Following Ref.~\cite{coulombe2017computing}, the amplitude-modulated speech signal $u(t)$ is applied everywhere, with a force $f_i(t)=a_0\alpha_i\sin(\omega_mt)|u(t)|$. At every site, the nonlinear parameter $\lambda_i$ is randomly chosen to be either $1$ or $4$, and $\alpha_i$ is chosen to be either $1$ or $7$. The modulation frequency is set to $\omega_m=1.085\omega_0$ and site frequency is $\omega_0/(2\pi)=3580Hz$ in speech file time.

We embody the model in a chain of resonating drums, mapping every degree of freedom to the first mode of vibration of each drum (Fig.~\ref{fig:speech}a). To encode the variable nonlinear parameter $\lambda_i$ in the geometry, we use two different arm designs: A short, straight arm, that presents high nonlinearity (because small displacements are significant in comparison with the arm length), and a broad, C-shaped arm with a longer effective length that remains linear at larger drum displacements.  We use gradient descent to identify the thickness of the beams that produces the same linear stiffness while achieving the necessary contrasting nonlinearities (Fig.~\ref{fig:speech}b). Although in this example we only engineer the local nonlinearity, it is also possible to control the coupling nonlinearity by engineering the connecting beam geometry (see Supplementary Information).

We simulate the corresponding tight-binding model in the binary classification task of distinguishing between utterances of the words yes and no from the TI-46 dataset (Fig.~\ref{fig:speech}c and Methods). These results demonstrate the need for nonlinearity in physical machine learning tasks: When the signal amplitude is too small, the system is linear and the classification error is fixed at $18\%$. However, as the amplitude is increased, the error reduces six-fold, reaching a minimum of $3\%$. Although here we have only optimized the readout function, the differentiability of the geometry allows for next-generation reservoir computing designs where the dynamical system is optimized together with the readout function.

\section*{Discussion and outlook}
Our results have shown that a variety of computational paradigms can be embodied in nonlinear perturbative metamaterials, by designing metamaterial geometries according to suitable functional tight-binding models. The method is limited to geometries where the spectral gap between modes of interest and higher modes is greater than the coupling strengths in the tight-binding model---a requirement that already exists in linear perturbative metamaterials~\cite{Matlack2018}, and an additional requirement of time scale separation between the metamaterial modes that compose the effective theory, and the modes driven by nonlinear forces. While these limitation means that the coordinate transformation in Eq.~\ref{eq:NonlinearCoordTransformation} may not be suitable for extracting tight-binding models from arbitrary metamaterial geometries, it does not preclude its use in design: when designing, one can exploit the freedom to chose a base geometry where the modeling assumptions are satisfied. Remarkably, the requirement for narrowband operation, which was assumed to be essential in prior works, is actually not strict, as we have shown by implementing a persistent memory based on multistable, quasistatic dynamics.

From a computational perspective, the tight-binding model extraction that we have introduced is very efficient: our algorithm has linear time complexity, and can run fully in parallel for every site. These properties, together with the additive and local relation between model and geometry, result in a very well posed design problem that can be scaled to large systems. We anticipate that this efficient design method---together with advances in nanoscale metamaterial integration~\cite{dorn2025graded}---will enable unprecedented information-processing functionalities in metamaterials. Nonlinear perturbative metamaterials also provide an avenue for the realization of analogs of interacting quantum phenomena in metamaterial systems~\cite{yuan2017creating, jurgensen2023quantized, wang2024realization}.

\section*{Acknowledgements}
We thank Tena Dub\v{c}ek and Oded Zilberberg for stimulating conversations. We would like to acknowledge Jorgen Dokker, Garth Wells and everyone else in the FEniCS community for their outstanding contribution to open science and the patient and helpful advice with the software. Figures ~\ref{fig:concept}, \ref{fig:cim}a, \ref{fig:topologicalSoliton}a and \ref{fig:speech}a have been produced by Laura Canil from Canil Visuals.

Funded by the European Union. Views and opinions expressed are however those of the author(s) only and do not necessarily reflect those of the European Union or the European Research Council Executive Agency. Neither the European Union nor the granting authority can be held responsible for them.

This work is supported by the ERC grant 101040117 (INFOPASS).

\bibliography{references}

\section*{Methods}

\subsection{Material model, parameters and finite element tools}
We model the material as a Kirchhoff-Saint Venant hyperelastic material model, with the Lagrangian density

\begin{equation}
\mathscr{L}=1/2\rho\dot{u}^2-\frac{\lambda}{2}\mathrm{Tr}^2\:(E)-\mu \mathrm{Tr}\:(E^2).
\label{eq:SysLag}
\end{equation}
Where $\rho$ is the density, $\lambda$ and $\mu$ are the Lam\'e coefficients, and $E=\frac{1}{2}(F^TF-I)$ is the Lagrange strain tensor, with $F=\nabla \vec{u}+\mathbb{I}$ representing the deformation gradient. Throughout the model, we set $\rho=1$, and express the Lam\'e coefficients $\lambda$ and $\mu$ in terms of a Young's modulus ($E=100$) and Poisson ratio ($\nu=0.33$). This is without loss of generality, as it is always possible to re-scale the Young's modulus and density by re-scaling the reduced mass and stiffness matrices, or equivalently changing the displacement, time and force units. This is done in Fig.~\ref{fig:speech} to match the resonator frequency to the speech recordings. The plotted relative accuracies remain valid for the re-scaled forces and displacements. Damping is not part of the finite element model, but is added after discretization by applying a velocity-dependent force proportional to the mass matrix, with a coefficient of proportionality of $0.05$ in Fig.~\ref{fig:NonlinearOscillation}, $0.001$ in Fig.~\ref{fig:cim}---lower than the baseline case, to allow for the mode corresponding to the solution of the Ising problem to be spectrally separated from the rest, $4.0$ in Fig.~\ref{fig:topologicalSoliton}---higher than the baseline case, as the system is meant to operate in an overdamped regime. In Fig.~\ref{fig:speech} a site quality factor of $60$ is specified, to match the model in the reference publication~\cite{coulombe2017computing}.

Finite element calculations are performed using the open source package FEniCSx~\cite{Scroggs2022, 10.1145/2566630, baratta_2023_10447666, 10.1145/3524456}. The geometry is defined in GMSH using the OpenCascade kernel, and a different mesh is constructed for every component of the model (see the methods section on handling modular geometries).

\subsection{Determination of the effective model}
The determination of the effective model is divided into two steps. In the first step, we determine the basis functions $\Psi_i$, which will be mapped to each tight-binding degree of freedom. In the second step, we extract the tight-binding parameters from the energy density (Eq.~\ref{eq:SysLag}), the basis functions and the geometry. Throughout this section and the Supplementary Information, we will refer to the basis functions as Wannier functions, as the metamaterials that we are considering can be seen as periodic system with a small disorder. But the same principle applies to heterogeneous models where the basis functions would be analogous to Localized Molecular Orbitals (LMO) in Density Functional Theory (DFT).

\subsubsection{Determination of the basis functions}

\begin{figure}[b]
    \centering
    \includegraphics[width=\linewidth]{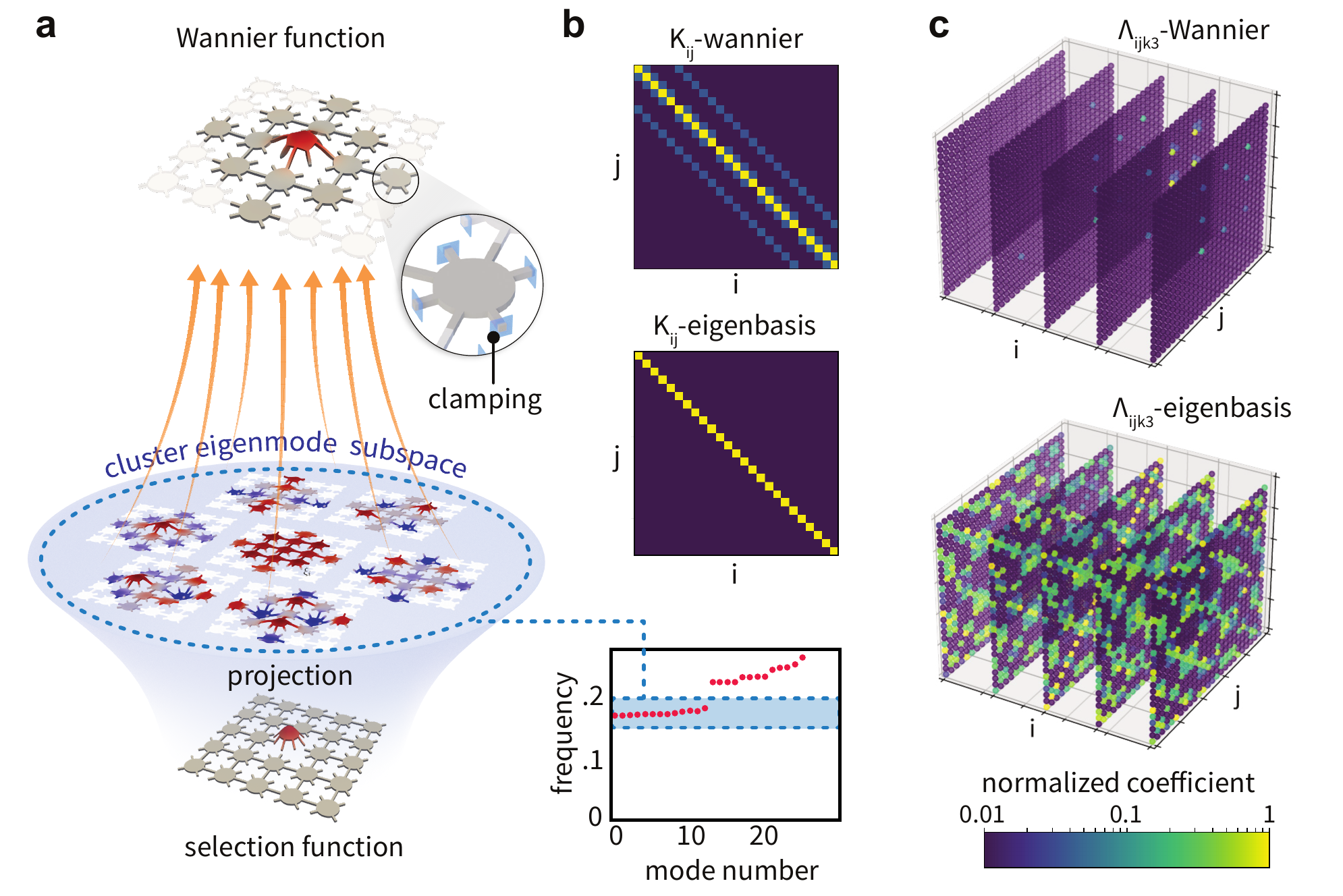} 
    \caption{\textbf{ | Modelling nonlinear perturbative metamaterials with maximally localized functions} \textbf{a} The Wannier functions are obtained by projecting a set of local symmetry-selection functions in the spectral subspace spanned by the cluster eigenmodes that fall within a frequency range of interest.
	 \textbf{b} In the Wannier function basis, the stiffness matrix $K_{ij}$ is banded, with the dominant off-diagonal terms induced by the geometric elements (beams) connecting the drums. This is in contrast with the stiffness matrix in the eigenstate basis, that is diagonal. \textbf{c} In the Wannier function basis, the Kerr energy term $\Lambda_{ijkl}$ is sparse, as it contains only interactions between nearby sites---these terms can be associated to geometric features, such as coupling beams, that can be combined to design objective modes. In contrast, in the eigenbasis, the Kerr energy term is dense. Its elements cannot be associated to geometric features, and the memory and time required to perform time-domain simulations scales quartically with system size.}
    \label{fig:figureMatrices}
\end{figure}

To compute the localized basis functions (Wannier functions) $\Psi_i(x)$ in Eq.~\ref{eq:NonlinearCoordTransformation}, we project a local symmetry-selection function $\xi_i(x)$---that identifies the site and orbital corresponding to the Wannier function---on the spectral subspace spanned by the set of modes in the frequency range of interest (Fig.~\ref{fig:figureMatrices}). Since the Wannier functions and Wannier derivatives decay rapidly, they can be approximated by a compact-support function, allowing us to perform computations on a fixed-size cluster around each site (See supplemetary information for extensive details on the implementation). This results in an efficient algorithm where the required computational time scales linearly with the number of sites in the metamaterial~\cite{goedecker1999linear}---enabling us to tackle the large number of sites required for information processing applications. 

The Wannier function sensitivities or Wannier derivatives $\Psi'(x)$ are computed by introducing a parameterized displacement in the finite element formulation, $u(x)=q_i\Psi_i$ and then computing the change in tangent stiffness matrix $dK/dq_i$ using the automatic differentiation feature built in UFL. Then, we identify the change in basis function $\Psi'=d\Psi_i/dq_i$ using eigenvector perturbation theory ~\cite{nelsonmethod}. It is not necessary to determine the full $dK/dq_i$, as the eigenvector sensitivity depends only on specific contractions $(dK/dq_i) \cdot v$ (see Supplementary Information for details on an efficient implementation). The resulting Wannier derivatives $\Psi'(x)$ can be interpreted as capturing the response of all the high-frequency modes, resulting from the residual nonlinear forces (See Supplementary Information and Ref.~\cite{rutzmoser2017generalization}). High-frequency modes displace to minimize, or `screen', regions of localized stress induced by the nonlinearity, and thus have the effect of reducing the effective nonlinear coefficients from their bare values.  

\subsubsection{Determination of the effective theory}
To extract the tight-binding model, we express the finite element displacement $u(x)$ in terms of tight-binding coordinates $q_i$, using Eq.~\ref{eq:NonlinearCoordTransformation}. Then, we Taylor-expand the energy, given by the volume integral of $\mathscr{L}(u)$ (Eq.~\ref{eq:SysLag}) in terms of $q_i$, using UFL built-in autodifferentiation capabilities. This yields the tight-binding model characterized by the energy
\begin{equation}
V(q)=\frac{1}{2}K_{ij}q_iq_j + \frac{1}{3}\Gamma_{ijk}q_iq_jq_k+ \frac{1}{4}\Lambda_{ijkl}q_iq_jq_kq_l,
\label{eq:EnergySeries}
\end{equation}  
with $K$ being the linear stiffness matrix, $\Gamma$ the second-order nonlinearity tensor, and $\Lambda$ the third-order (Kerr or Duffing) nonlinearity tensor. These tensors are approximately sparse (Fig.~\ref{fig:figureMatrices}b,c), where the dominant terms are associated to identifiable geometric features such as coupling beams between site and neighbor. This crucial requirement for design is met by the use of Wannier functions as basis. If a basis of eigenstates is used, the resulting problem presents dense nonlinear interactions~\cite{JAIN201780}, (Fig.~\ref{fig:figureMatrices}b,c)---each of them depending on all geometric features, that consequently cannot be easily mapped into the objective tight-binding model.

Each coefficient is obtained by summing the energy contributions from the sites where the basis functions corresponding to all relevant degrees of freedom are nonzero, i.e., the sites where the all clusters involved in the computation overlap. For example, the term $K_{ij}$ is evaluated only in the sites lying at the intersection of the clusters on which $\Psi_i$ and $\Psi_j$ are defined.


\subsubsection{Accuracy and validity of the model}
\begin{figure}[b!]
	\centering
	\includegraphics[width=1.0\linewidth]{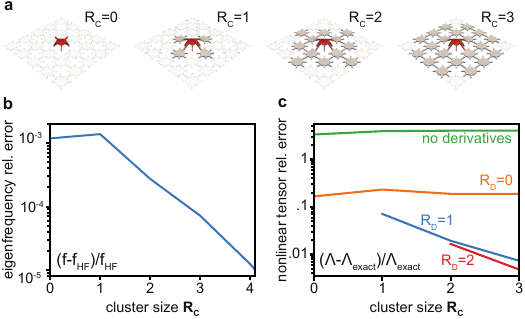}
	\caption{\textbf{| Accuracy of the computation as a function of cluster size and Wannier derivative computation radius} \textbf{a} Wannier functions computed with different number of nearest neighbours, controlled by the radius (R) parameter. R=0 indicates that only one site is used. \textbf{b} Relative error of the linear eigenfrequencies as a function of the cluster size. \textbf{c} Accuracy of the nonlinear tensor, defined as $|\Lambda-\Lambda_{exact}|/|\Lambda_{exact}|$. Different lines are calculated by including a different number of basis function sensitivities. For the green line, no sensitivities are used, and the nonlinear tensor is overestimated by a factor of roughly 4. For the orange line, only the sensitivity of each basis function with respect to itself is used. In the blue (red) line, up to nearest (next nearest) neighbours are included. 
	\label{fig:clusters}}
\end{figure}
The accuracy of the extracted tight-binding model depends on the size of the cluster used to determine the Wannier functions (Fig.~\ref{fig:clusters}b,c). Here we fix the cluster size though a topological distance cutoff; with a radius $R_C=0$, only the site associated to the Wannier function is used in the Wannier function calculation, with $R_C=1$, the cluster includes nearest-neighbors, and with $R_C=2$, next nearest neighbors. The connections between the cluster sites and neighbors are left as open boundary conditions. Since the Wannier functions are exponentially localized, the model parameters converge very quickly to the correct value. For every site, we include only a few Wannier derivatives  $\Psi'_{ij}$, corresponding to pairs of sites that are within a topological distance $d(i,j) \leq R_D$. Excluding all derivatives essentially eliminates the nonlinear contribution to the map in Eq.~\ref{eq:NonlinearCoordTransformation}, and results in bare nonlinear parameters that deviate significantly from the correct valye. For example, the Kerr (Duffing) coefficient is off by a factor of 4 (Fig.~\ref{fig:clusters}c). Including only one Wannier derivative per site produces a renormalized value that already reduces this error by a factor of approximately 20 (Fig.~\ref{fig:clusters}c). 

The approximation in Eq.\ref{eq:NonlinearCoordTransformation} is only valid when there is a separation of time scales~\cite{JAIN201780}, because the modes in $\Psi'$ are assumed to respond instantaneously to changes in $q$. This is the case in the thin elastic metasurfaces considered here, where the nonlinearity induces tension changes, and these propagate much faster than the flexural oscillations of the resonators. In such systems, the nonlinear coordinate transformation is analogous to the Born-Oppenheimer approximation~\cite{bornop} in quantum chemistry, where electrons are assumed to respond instantaneously to changes in the positions of the nuclei---here, in-plane modes captured by $\Psi_{ij}'$ play the role of the electrons, and flexural modes in $\Psi_i$ play the role of the nucleii. In general, one can use the flexibility in choosing the metamaterial geometry so that the time scale separation assumption remains valid, following the design-for-analysis principle~\cite{suri1989design}. 

The approach used in this work differs from that of the original perturbative metamaterials framework developed by Matlack et al.~\cite{Matlack2018} In the original work, an intermediate effective theory, incorporating a large number of local modes, is first constructed on a two-site cluster. Then, the actual effective theory is distilled from this high dimensional model via a Schrieffer-Wolff transformation. We have observed that directly building the final effective theory using larger clusters provides better numerical stability in systems with lower spectral gaps.

\subsection{Handling modular geometries}

Complex information processing requires a large number of degrees of freedom---be it parameters on a Large Language Model (LLM), or transistors in a CPU. The design algorithms introduced in this Article are particularly suited to such large scale problems, because the time complexity of the method is linear with the number of sites: Basis functions and model parameters are computed in fixed clusters around each site, and the time necessary to solve for a cluster is constant. The memory complexity is also benign: It is possible to operate in constant memory, if clusters are solved sequentially and not cached, although parallel algorithms and caching (of cluster eigenmodes, factorized site and cluster matrices, etc.) will improve the speed at the expense of increased memory requirements. Taking advantage of the benign complexity of our design method requires carefully handling the modular geometry of the metamaterial. 
\begin{figure}[t!]
	\centering
    \includegraphics[width=\columnwidth]{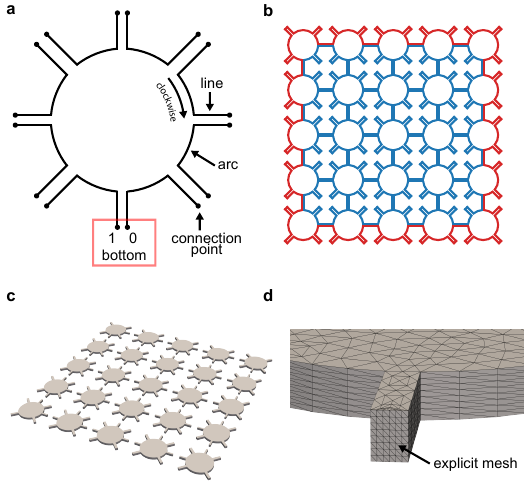}
\caption{ \textbf{| Handling modular geometries} 
{\fontfamily{phv}\selectfont} Modular geometries are extruded from planar designs, composed of individual components. Each polyline of the geometry is constructed in clockwise order. This allows the system to determine the inside and outside of the structure. \textbf{b} A constraint enforcement system constructs the whole geometry, adjusting the parameters to match the connection points between connected components.  \textbf{c} Each component is meshed individually. This allows for parallelism or low memory operation, as components can be simulated sequentially or in parallel.  \textbf{d} To allow the simulation of the full system, finite element meshes at the connection points are explicitly meshed.
The color (red or blue) indicate positive (clockwise) or holes (counterclockwise) }
\label{fig:modularGeometry}
\end{figure}

We represent the geometry as a graph of components. In all examples except Fig.~\ref{fig:topologicalSoliton}e, components correspond to metamaterial sites. In Fig.~\ref{fig:topologicalSoliton}e, we use separate components for the beam and the coupling elements, to facilitate the geometry assembly. The geometry of every component is represented as a set of \emph{polylines}, which consist of three types of elements: Connection points, lines and circle arcs. Lines and arcs directly translate to geometric features of the final designs; in contrast, connection points, indicate where a component will connect to the next one, or where a boundary condition will be placed. The data structures representing the polylines are lightweight, storing only a few numerical values (two for the connection point, four for the line, and 6 for the circle arc), and are generated on the fly from a set of parameters. In addition, connection points can be grouped and named---providing a way to indicate that multiple connection points represent the same logical boundary.

To input a design, the user provides a Python function that generates a set of polylines from a set of parameters. By using JAX to generate the geometry from the parameters, the design is diffferentiable, even if the features are determined by a complex function of the geometric parameters. Figure~\ref{fig:modularGeometry}a shows the geometry of a drum from Fig~\ref{fig:NonlinearOscillation}. The geometry contains 8 polylines, each of which consisting of a connection point, a line segment, a circle arc, a second line segment, and a second connection point. These parameters are specified in a clockwise order, observed from the inside of the material component. This allows the code to automatically determine which parts of the design are filled with material. To complete the metamaterial design, the user provides the connectivity between components. This is indicated by providing a list of connections, that indicate that a named group of connection points from one component is connected to a different named group of connection points in a different component---for example, in a chain, the \emph{right} group of connection points on \emph{site\_1} will be connected to the \emph{left} group of connection points on \emph{site\_2}. Because the coordinates of the connection points are automatically differentiable, the code can lay out the geometry by finding the values of a set of parameters that minimize the difference between connected connection points, enabling the simple generation of complex layouts (Fig.~\ref{fig:modularGeometry}b). Although the code does not prevent collisions between geometric features, these can be avoided by adequately parameterizing the geometry---for example, by describing feature locations in terms of the gap sizes, and constraining the gaps to be positive. Determining the geometric parameters that satisfy the connection constraints does not require meshing the geometry, and is as such a lightweight operation. Once the constraint values have been identified, the code also provides the Jacobian of the distance of every pair of connected connection points as a function of the geometric parameters---which can be used to ensure that the geometry remains connected during optimization.

Once the geometric parameters have been determined, the components can be meshed individually. We do so by transferring the geometry to GMSH using the OpenCascade geometry kernel. The geometry is then extruded by the corresponding device thickness (Fig.~\ref{fig:modularGeometry}c) and meshed. The meshes at every pair of connection points are defined explicitly, ensuring that the degrees of freedom in connected components conform (Fig.~\ref{fig:modularGeometry}d). This work is concerned with planar geometries (such as those that can be microfabricated in a silicon wafer), and it is straightforward to extend the code to multi-layer devices (e.g. involving a thin film of $Si_3N_4$ on a silicon device).

\subsection{High-fidelity simulation}
In Figs~\ref{fig:NonlinearOscillation} and~\ref{fig:cim}, we perform full-wave, nonlinear high fidelity simulations using the Generalized Alpha algorithm, in the form described by Arnold and Bruls~\cite{arnold2007convergence}. The time step is $h=T/192$ for Fig.~\ref{fig:NonlinearOscillation}, $h=T/628$ for Fig.~\ref{fig:cim}---the reason for using a smaller time step in Fig.~\ref{fig:cim} is that, in this case, the quality factor is higher, and hence the results are more sensitive to numerical errors. This high quality factor is due to the fact that the resonance peaks of all modes in the CIM are close together, but the system must exhibit a higher differential response for the mode corresponding to the correct solution. 

The high-fidelity results in Fig.~\ref{fig:topologicalSoliton} are computed using equilibrium continuation, as parameters are varied quasistatically. At every simulation step, we perturb the boundary conditions by increasing the pump phase by $2\pi/100$. Then, we identify the new equilibrium deformation using a conjugate gradient algorithm with line search to minimize the energy. In Fig.~\ref{fig:speech}, no high-fidelity simulations can be conducted, as the size of the problem precludes it. However, the regime at which the maximum classification accuracy is achieved is close to that in Fig.~\ref{fig:NonlinearOscillation}d in terms of relative nonlinearity. Thus the model is expected to be highly accurate in this regime.

\subsection{Tight-binding simulations}
 
In Figs.~\ref{fig:NonlinearOscillation} and \ref{fig:topologicalSoliton}, we perform the time-domain tight-binding simulations using integrate.solve\_ivp function from the Python package scipy with default parameters. In Figure ~\ref{fig:speech} , we use the Dormand-Prince solver with order 8/7 from the  package Diffrax. The step size is determined by the built-in PID controller, with an absolute tolerance of $10^{-7}$ and a relative tolerance of $10^{-16}$. The use of Diffrax allows us to compile the integration loop, with a resulting increase in simulation efficiency. This is necessary to tackle the comparatively larger problem size. In Fig.~\ref{fig:topologicalSoliton}, we also perform equilibrium continuation with a time step $2\pi/100$.

\subsection{Training of the speech classification model}

To train the speech classification system, we simulated the response of the system using utterances from 8 female speakers from the TI-46 dataset, following the test-training split provided by the dataset. From this simulation, we identify the mean vibration amplitude at every site by computing
\begin{equation}
A_i=\sum_n x_i\sin(\omega_mt).
\end{equation}

The weights are then normalized according to the mean amplitude of the training set.

Following Ref.~\cite{coulombe2017computing}, we divide the chain into overlapping segments. For every overlapping segment, we train a classifier using the Fisher discriminant procedure with Tikhonov regularization, obtaining the weights according to the equation:
\begin{equation}
w_i = (S+\sigma I)^{-1}(m_2-m_1),
\end{equation}

where $m_1$, $m_2$ are the feature means computed for every class, and $S$ is the within-class scattering matrix, computed as $\sum_{i\in classes}(X_i-m_i)^T(X_i-m_i)$. The regularization parameter $\sigma$ is set to $10^{-12}$, and we empirically confirm that changing it by an order of magnitude does not affect the results. Then, we perform inference on every segment $s_i$ by combining the vibration amplitudes at each site according to the weights $w_i$ ($1<=i<=n$) determined from the Fisher discriminant,
\begin{equation}
o_i=\sum_{j\in s_i}{w_iE_i}.
\end{equation}
The final result is obtained by performing majority inference on the outputs of all segments.





\appendix

\newcommand{\myfrac}[3][0pt]{\genfrac{}{}{}{}{\raisebox{#1}{$#2$}}{\raisebox{-#1}{$#3$}}}

\section{Designing nonlinear interactions}
\begin{figure}[b!]
	\centering
	\includegraphics[width=1.0\linewidth]{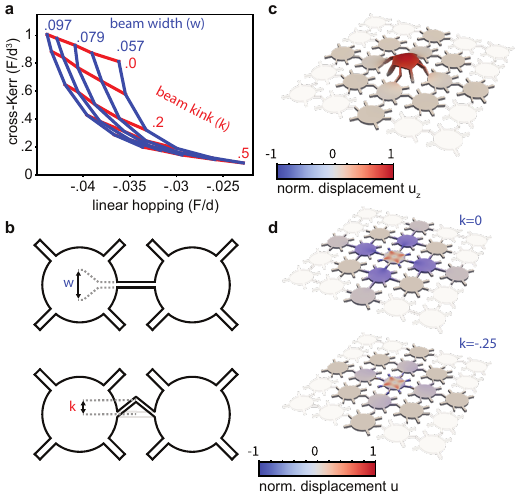} 
    \caption{Controlling nonlinear interactions though coupling geometry | \textbf{a} Linear coupling coefficient and cross-Kerr interaction strength as a function of the beam geometric parameters. We can independently control the coupling stiffness and cross-Kerr nonlinearity.  \textbf{b} The parameters $k$ ald $w$ define the geometry of the connecting beam. These are used to independently control the cross-Kerr interaction and the linear coupling. \textbf{c} Wannier function corresponding to a vibration localized at the center site. \textbf{d} Wannier derivatives corresponding to geometries with different kink value $k$.} 
    \label{fig:nonlinearControl}
\end{figure}

In the speech classification example, we engineered the local nonlinear interaction by designing the support geometry. A similar mechanism allows us to design coupling nonlinearities. Here, we illustrate this by designing the cross-Kerr interaction between neighboring drums (Fig.~\ref{fig:nonlinearControl}a). Cross-Kerr nonlinearities are represented by potential terms of the form $q_i^2q_j^2$, and cause each site to experience a frequency shift proportional to the displacement squared (approximately, the energy) in the other site. 

The cross-Kerr nonlinearity is controlled by inducing a kink in the connecting beam (Fig.~\ref{fig:nonlinearControl}b). Straight beams result in a highly nonlinear coupling, because they are efficient at transmitting vibration-induced tension changes. In contrast, kinked beams can relax this horizontal tension and mitigate cross-Kerr interactions. This can be corroborated by looking at the modal derivatives. When a given site is vibrating according to its Wannier function (Fig.~\ref{fig:nonlinearControl}c), the basis function derivative is highly delocalized when the beams are straight (Fig.~\ref{fig:nonlinearControl}d), indicating a strong transfer of in-plane force to the neighbors. In contrast, when the connecting beams are kinked, the modal derivative becomes delocalized.

Remarkably, it is possible to independently control the linear and nonlinear coupling coefficient (Fig.~\ref{fig:nonlinearControl}a), by modifying two independent geometric parameters of the coupling beam (Fig.~\ref{fig:nonlinearControl}b).

\section{Determining the equations of motion for the effective theory}

This section presents a derivation of the tensor coefficients in Eq.~\ref{eq:EnergySeries} from the derivatives of the Lagrangian density (Eq.~\ref{eq:SysLag}). We start from the map between tight-binding degrees of freedom $q_{i}(t)$ and the metamaterial deformation $u(x, t)$ (Eq.~\ref{eq:NonlinearCoordTransformation}), repeated here for convenience:

\begin{equation}
u_{\mu}(x,t) = q_i(t)\Psi_{\mu i}(x)+\frac{1}{2}q_i(t)q_j(t)\Psi_{\mu ij}'(x).
\label{eq:mapping_equ}
\end{equation}
For simplicity, in this work all equations are expressed using Einstein summation convention. Indices $\mu$, and $\nu$ are reserved for the full model (FEM model), while $i$, $j$, $k$ and $l$ are allocated for the reduced (tight-binding) model. In the above equation, the displacement field $u$ has dimension $1 \times n$, where $n$ represents the degree of freedom of the full model (FEM model). The matrix $\Psi$ consists of all Wannier functions, designated as basis vectors. $\Psi$ has dimension $n\times n^r$. Where, $n^r$ denotes the degree of freedom of the tight-binding model. The superscript $r$ is assigned to the tight-binding (reduced) model. The tensor $\Psi'$ represents the sensitivity of the Wannier functions with respect to displacements of the tight-binding degrees of freedom. The variable $q$ denotes the tight-binding degree of freedom, here referred to as the Wannier coordinate.

The Wannier functions $\Psi_{\mu i}$ are linear combinations of cluster eigenvectors that maximize the vibration at specific sites,
\begin{equation}
\Psi_{\mu (i)} = \Phi_{\mu j} \tau_{(i)j},   
\label{WF}
\end{equation}
where $\Phi_{\mu j}$ is the $j^{th}$ cluster eigenmode, and $\tau{ij}$ is the linear combination coefficient. These coefficients are obtained by projecting a symmetry selection function $\xi_{\nu (i)}$ into the cluster eigenmodes
\begin{equation}
\hat{\tau}_{(i)j}=\Phi_{\mu j}M_{\mu \nu}\xi_{\nu (i)},
\label{eqn:WannierCoefficientsHat}
\end{equation}
where $M_{\mu \nu}$ is the cluster mass matrix obtained from finite element simulation. The function $\xi_{\nu (i)}$ identifies which Wannier function is going to be calculated. It is only nonzero in the associated site, and has a symmetry characteristic that corresponds to the required site orbital. Figure~\ref{fig:figureMatrices}a shows a symmetry selection function $\xi$ and the resulting Wannier function. From $\hat{\tau}_{(i)j}$, the coefficients are computed via the normalization
\begin{equation}
\tau_{(i)j}=\myfrac[3pt]{\hat{\tau}_{(i)j}}{\sqrt{\hat{\tau}_{(i)k}\hat{\tau}_{(i)k}}}.
\label{eqn:WannierCoefficients}
\end{equation}
This normalization condition is obtained from the mass-norm of the Wannier function, $\left|\Psi_{(i)}\right|=\sqrt{\Psi_{\mu (i)}M_{\mu\nu}\Psi_{\nu (i)}}$, by combining Eq.~\ref{eqn:WannierCoefficientsHat}  and Eq.~\ref{WF}, and using the fact that the normal modes $\Phi_{\mu (i)}$ are mass-orthonormal, i.e. $\Phi_{\mu i}\Phi_{\mu j}=\delta_{ij}$. 
To derive the equation of motion, the Euler-Lagrange equation is utilized, with extended terms incorporated to account for energy dissipation due to viscous friction, as follows:
\begin{equation}
\myfrac[0pt]{d}{dt}\myfrac[3pt]{\partial L}{\partial \dot{q}}-\myfrac[3pt]{\partial L}{\partial q} + \myfrac[3pt]{\partial R}{\partial \dot q} = 0
\end{equation}
The Lagrangian function, $L$, is defined as $L = T - U$, where $U$ defines elastic potential energy and $T$ is kinetic energy. The term $R$ accounts for energy dissipation due to viscous friction. External forces can be embedded into the potential energy form as:
\begin{equation}
V = \int u_{\mu} \cdot f^{s}_{\mu}(x)f^{t}_{\mu}(t)dx
\end{equation}

where the external force is decoupled into spatial $f^{s}_{\mu}$ and temporal $f^{t}_{\mu}$ components. Since partial derivatives are linear operations, we can treat the elastic, kinetic and energy dissipation parts of the Euler-Lagrange equation separately, as follows: 
\begin{equation}
\left [\myfrac[3pt]{d}{dt}\myfrac[3pt]{\partial T}{\partial \dot{q}}-\myfrac[3pt]{\partial T}{\partial q}\right] - \left [\myfrac[3pt]{d}{dt}\myfrac[3pt]{\partial U}{\partial \dot{q}}-\myfrac[3pt]{\partial U}{\partial q}\right] + \myfrac[3pt]{\partial R}{\partial \dot q} = 0
\label{Euler_L2}
\end{equation}

\textbf{Kinetic terms:} The first terms are related to the kinetic energy, which is defined as:
\begin{equation}
T = \frac{1}{2} \dot{u}_\mu  M_{\mu \nu}  \dot{u}_\nu
\label{kinetic_energy}
\end{equation}
here $M$ represents the mass matrix derived from the finite element method. By utilizing the equation \ref{eq:mapping_equ}, the velocity vector of the full model, $\dot u$ is given by:
\begin{equation}
\begin{array}{r@{\;}l@{\;}l@{\;}}
\dot{u}_\mu& =& \Psi_{\mu i}\dot{q}_i + \frac{1}{2} \Psi_{\mu i j}'\dot{q}_i q_j + \frac{1}{2}\Psi_{\mu i j}'q_i \dot{q}_j \\[11pt]
\end{array}
\end{equation}
It should be noted that, in Einstein summation, the indices may be swapped. To facilitate the rewriting of the above equation, $\hat{\Psi}_{\mu i j}$ is introduced:

\begin{equation}
\hat{\Psi}_{\mu i j} = \frac{1}{2} \left ( \Psi_{\mu i j}'+ \Psi_{\mu j i }' \right )    
\end{equation}
 While $\Psi'$ is not always symmetric (e.g., when static modal derivatives are not employed), $\hat{\Psi}$ is always symmetric. It should be noted that, $\hat{\Psi}$ consists of the derivative of the basis vectors. The equation above can be simplified as follows:
\begin{equation}
\begin{array}{r@{\;}l@{\;}l@{\;}}
\dot{u}_\mu& =& \left [ \Psi_{\mu i} + \hat{\Psi}_{\mu i j} q_j \right] \dot{q}_i
\label{u_dot}
\end{array}
\end{equation}

by substituting the above equation into the equation \ref{kinetic_energy}, the the kinetic energy can be expressed as:

\begin{equation}
T = \frac{1}{2} \dot{q}_i \left [ \Psi_{\mu i} + \hat{\Psi}_{\mu i j} q_j \right]  M_{\mu \nu} \left [ \Psi_{\nu k} + \hat{\Psi}_{\nu k l} q_l \right]  \dot{q}_k
\end{equation}

Using the kinetic energy in Euler-Lagrange equation and rewriting it leads to the following results:

\begin{equation}
\begin{array}{r@{\;}l@{\;}l@{\;}}
\myfrac[3pt]{d}{dt}\myfrac[3pt]{\partial T}{\partial \dot{q}}-\myfrac[3pt]{\partial T}{\partial q} &=& \underline{\Psi_{\mu i} M_{\mu \nu} \hat{\Psi}_{\nu k l} \dot{q}_k \dot{q}_l} +\\[11pt]
&& \Psi_{\mu i} M_{\mu \nu} \Psi_{\nu k} \ddot{q}_k +\\[11pt]
&& \underline{\Psi_{\mu i} M_{\mu \nu} \hat{\Psi}_{\nu k l} q_l \ddot{q}_k } +\\[11pt]
&& \underline{\underline{\hat{\Psi}_{\mu i j} q_j M_{\mu \nu} \hat{\Psi}_{\nu k l} \dot{q}_k \dot{q}_l }} +\\[11pt]
&& \underline{\hat{\Psi}_{\mu i j} q_j M_{\mu \nu} \Psi_{\nu k} \ddot{q}_k} +\\[11pt]
&&\underline{\underline{\hat{\Psi}_{\mu i j} q_j  M_{\mu \nu} \hat{\Psi}_{\nu k l} q_l \ddot{q}_k}} +
\end{array}
\end{equation}

The second-order terms are denoted by a single underline, while the third-order terms are represented by a double underline. When the modal derivatives are mass-orthogonal to the modes, all second-order terms (underlined) vanish. After applying these conditions and reordering the expression, the resulting equation is:

\begin{equation}
\begin{array}{r@{\;}l@{\;}l@{\;}}
\myfrac[3pt]{d}{dt}\myfrac[3pt]{\partial T}{\partial \dot{q}}-\myfrac[3pt]{\partial T}{\partial q} &=& \left ( M_{i k}^r +  \Lambda_{ijkl}^{mr} q_j q_l \right )\ddot q_k + \\[11pt]
&& \Lambda_{ijkl}^{mr} q_j \dot q_k \dot q_l 
\label{kinetic_t}
\end{array}
\end{equation}

Based on the above equation, the reduced mass matrix $M^r$ and the rank-four inertia tensor $\Lambda^{mr}$ are determined as follows:
\begin{equation}
    M^r_{ik} = \hat{\Psi}_{\mu i} M_{\mu \nu} \hat{\Psi}_{\nu k}
    \label{reduced_mass}
\end{equation}
\begin{equation}
    \Lambda_{ijkl}^{mr} = \hat{\Psi}_{\mu i j}  M_{\mu \nu} \hat{\Psi}_{\nu k l} 
    \label{reduced_inirtia_lambda}
\end{equation}

The $M$ matrix derived from the FEM model and has dimension $n \times n$, in contrast, $M^r$ represents the reduced mass matrix with dimension $n^r \times n^r$. The relationship $M^r = I$ holds when the eigenvectors are mass normalized. The additional tensor $\Lambda^{mr}$ appears in the preceding equation due to the nonlinearity of the coordinate transformation in Eq.~\ref{eq:NonlinearCoordTransformation}. The nonlinearity defines a non-inertial frame of reference, resulting in additional force terms. A similar reasoning will be applied later to the damping matrix. \\

\textbf{Elastic term:} The second part is related to the elastic potential energy. The energy in the WF reduced model is assumed to take the following form:
\begin{equation}
\begin{array}{r@{\;}l@{\;}l@{\;}}
U(q) &=& \frac{1}{2} q_i K_{ij}^r q_j + \frac{1}{3} q_i \Gamma_{ijk}^r q_j q_k +\\[11pt]
&& \frac{1}{4} q_i \Lambda_{ijkl}^r q_j q_k q_l + q_i f_i^{s^r}
\label{potential_eng}
\end{array}
\end{equation}

The matrices $K^r$ is the reduced linear stiffness matrix, while $\Gamma^r$ and $\Lambda^r$ are reduced non-linear stiffness matrices. The vector $f^{s^r}$ indicates the reduced spatial force vector, assuming $f^t = 1$. The reduced matrices and force vector can be derived by successively differentiating the above equation and setting $q=0$. 

\begin{equation}
\begin{array}{r@{\;}l@{\;}l@{\;}}
\myfrac[3pt]{\partial U(q)}{\partial q} \bigg 
| _{q=0} & = & \Bigl [ K_{ij}^r q_j + \Gamma_{ijk}^r q_j q_k + \Lambda_{ijkl}^r q_j q_k q_l + \\[11pt]
&& f_i^{s^r}\Bigr ] _{q=0} \\[11pt]
&=& f_i^{s^r}
\end{array}
\end{equation}
\begin{equation}
\begin{array}{r@{\;}l@{\;}l@{\;}}
\myfrac[3pt]{\partial^2 U(q)}{\partial q \partial q} \bigg 
| _{q=0} & = &\left [ K_{ij}^r + 2 \Gamma_{ijk}^r q_k + 3 \Lambda_{ijkl}^r q_k q_l \right ] _{q=0} \\[11pt]
& = & K_{ij}^r
\end{array}
\end{equation}

\begin{equation}
\begin{array}{r@{\;}l@{\;}l@{\;}}
\myfrac[3pt]{\partial^3 U(q)}{\partial q \partial q \partial q} \bigg 
| _{q=0} & = & \left [ 2 \Gamma_{ijk}^r + 6 \Lambda_{ijkl}^r q_l \right ] _{q=0} = 2 \Gamma_{ijk}^r
\end{array}
\end{equation}

\begin{equation}
\begin{array}{r@{\;}l@{\;}l@{\;}}
\myfrac[3pt]{\partial^4 U(q)}{\partial q \partial q \partial q \partial q} \bigg 
| _{q=0} & = & 6 \Lambda_{ijkl}^r
\end{array}
\end{equation}

After defining the reduced matrices with the aid of the above equations, the Euler-Lagrange equation can be formulated as follows:

\begin{equation}
\begin{array}{r@{\;}l@{\;}l@{\;}}
-\myfrac[3pt]{d}{dt}\myfrac[3pt]{\partial U}{\partial \dot{q}}+\myfrac[3pt]{\partial U}{\partial q} &=& K_{ij}^r q_j + \Gamma_{ijk}^r q_j q_k +\\[11pt] 
&&\Lambda_{ijkl}^r q_j q_k q_l + f_i^r
\label{elastic_t}
\end{array}
\end{equation}

\textbf{Damping term:} In the extended Euler-Lagrange equation the final term is associated with the damping effect.

\begin{equation}
\myfrac[0pt]{d}{dt}\myfrac[0pt]{\partial L}{\partial \dot{q}}-\myfrac[0pt]{\partial L}{\partial q} + \myfrac[0pt]{\partial R}{\partial \dot q} = 0
\end{equation}

Where $R$ corresponds the Rayleigh's dissipation, defined by the following equation:

\begin{equation}
    R = \frac{1}{2} \dot u_\mu B_{\mu \nu} \dot u_\nu
\end{equation}

The velocity vector $\dot u$ is defined in equation \ref{u_dot} and $B_{\mu\nu}$ represents the damping matrix extracted from FEM. Consequently, $R$ can be expressed as:

\begin{equation}
R = \frac{1}{2} \dot{q}_i \left [ \Psi_{\mu i} + \hat{\Psi}_{\mu i j} q_j \right]  B_{\mu \nu} \left [ \Psi_{\nu k} + \hat{\Psi}_{\nu k l} q_l \right]  \dot{q}_k
\end{equation}

To construct the final term in the Euler-Lagrangian equation, it is necessary to obtain the derivative of $R$ with respect to the velocity.

\begin{equation}
    \myfrac[3pt]{\partial R}{\partial \dot q} = \left [ \Psi_{\mu i} + \hat{\Psi}_{\mu i j} q_j \right]  B_{\mu \nu} \left [ \Psi_{\nu k} + \hat{\Psi}_{\nu k l} q_l \right]  \dot{q}_k
\end{equation}

Thus, the reduced damping matrix is given by:

\begin{equation}
    \myfrac[3pt]{\partial R}{\partial \dot q} = B^r_{ik} \dot{q}_k
\end{equation}

\begin{equation}
B^r_{ik} = \left [ \Psi_{\mu i} + \hat{\Psi}_{\mu i j} q_j \right]  B_{\mu \nu} \left [ \Psi_{\nu k} + \hat{\Psi}_{\nu k l} q_l \right] 
\end{equation}

If the damping matrix is defined as an inertia damping matrix, i.e., proportional to the mass matrix with Rayleigh coefficient $\alpha $ then: 

\begin{equation}
B_{\mu\nu} = \alpha M_{\mu \nu}
\end{equation}

\begin{equation}
\begin{array}{r@{\;}l@{\;}l@{\;}}
B^r_{ik} & = & \Psi_{\mu i} \alpha M_{\mu \nu} \Psi_{\nu k} + \underline{\Psi_{\mu i}\alpha M_{\mu \nu}\hat{\Psi}_{\nu k l} q_l } +\\ [11pt]
&&\underline {\hat{\Psi}_{\mu i j} q_j \alpha M_{\mu \nu} \Psi_{\nu k} } + \hat{\Psi}_{\mu i j} q_j \alpha M_{\mu \nu} \hat{\Psi}_{\nu k l} q_l 
\end{array}
\end{equation}

Under the assumption that the modal derivatives are mass-orthogonal to the modes, the underlined terms become zero:

\begin{equation}
B^r_{ik}  =  \Psi_{\mu i} \alpha M_{\mu \nu} \Psi_{\nu k} + \hat{\Psi}_{\mu i j} q_j \alpha M_{\mu \nu} \hat{\Psi}_{\nu k l} q_l 
\end{equation}

by subsisting equation \ref{reduced_mass} and \ref{reduced_inirtia_lambda}, the following equation for the damping term is obtained:

\begin{equation}
    \myfrac[3pt]{\partial R}{\partial \dot q} = \alpha \left ( M^r_{ik} +  \Lambda^{mr}_{ijkl} q_j  q_l \right) \dot{q}_k
    \label{damping_t}
\end{equation}


Finally, equations \ref{kinetic_t}, \ref{elastic_t} and \ref{damping_t} can be substituted into the equation \ref{Euler_L2} to obtain the complete form of Euler-Lagrangian equation: 

\begin{equation}
\begin{array}{r@{\;}l@{\;}l@{\;}}
&& \left ( M_{i k}^r + \Lambda_{ijkl}^{mr} q_j q_l \right )\ddot q_k + \Lambda_{ijkl}^{mr} q_j \dot q_k \dot q_l + K_{ij}^r q_j + \\[11pt] 
&&\Gamma_{ijk}^r q_j q_k  +  \Lambda_{ijkl}^r q_j q_k q_l + f_i^r + \\[11pt] 
&&\alpha \left ( M^r_{ik} +  \Lambda^{mr}_{ijkl} q_j  q_l \right) \dot{q}_k = 0
\label{reduced_eq_motion}
\end{array}
\end{equation}

\textbf{Assembling the equations of motion}

To solve the tight-binding model, we assemble a system of first order equations with dimension $2n^r$, where $n^r$ is the number of basis functions in the model. The resulting equations of motion are given by

\begin{equation}
\myfrac[3pt]{d q_k}{dt}=\dot q_k
\end{equation}
\begin{equation}
\begin{array}{r@{\;}l@{\;}l@{\;}}
\myfrac[3pt]{d \dot q_k}{dt} & = & - \left ( M_{i k}^r + \Lambda_{ijkl}^{mr} q_j q_l \right )^{-1} \Bigl [ \Lambda_{ijkl}^{mr} q_j \dot q_k \dot q_l +\\[11pt] 
&& K_{ij}^r q_j + \Gamma_{ijk}^r q_j q_k  +  \Lambda_{ijkl}^r q_j q_k q_l + f_i^r + \\[11pt] 
&& \alpha \left ( M^r_{ik} +  \Lambda^{mr}_{ijkl} q_j  q_l \right) \dot{q}_k \Bigr ]
\end{array}
\end{equation}

Once $q$ is calculated, the displacement vector $u$ of the full model can be obtained by using equation \ref{eq:mapping_equ}. Although the model involves a matrix inversion, this does not pose a significant challenge for the problem sizes considered here. Our observation is that the nonlinear mass terms are comparatively very small, indicating that, for large systems, significant performance gains can be achieved by Taylor-expanding the left hand side  $M_{i k}^r + \Lambda_{ijkl}^{mr} q_j q_l$.

\section{Numerical determination of the cluster eigenmodes}\label{sec:WannierComputation}
Calculating the Wannier functions requires finding the smallest eigenmodes of each cluster. We accomplish this by using the Implicitly-Restarted Arnoldi method of SLEPc together with a shift-and-invert spectral transform. The algorithm looks for the largest eigenpairs of the matrix $A=\left( K-\sigma M \right)^{-1}$, which correspond to the smallest eigenpairs of the original system. In our implementation, the matrix $A$ is never explicitly calculated, but instead defined as a matrix-vector multiplication.  Around zero shift ($\sigma=0$), this matrix-vector multiplication can be realized by solving a linear system with the cluster stiffness matrix as left-hand side.

The na\"ive approach to compute the spectral transform consists in factorizing the stiffness matrix for every cluster individually. This approach has the disadvantage of requiring the assembly and factorization of relatively large matrices, and does not take advantage of the fact that sites are shared among many clusters. To improve the efficiency of the methods, we use a substructuring approach. First, the stiffness matrix for every site is factorized independently. Then, the factorized stiffness matrix is used to compute the statically condensed site stiffness matrix, that is the force-displacement relation at the boundary degrees of freedom---these two calculations are amortized among all clusters that include the specific site. The statically condensed cluster stiffness matrix is calculated just before computing the cluster modes or cluster modal derivatives. This is done to reduce the memory footprint, as this is a dense matrix of significant size, and the re-computation time is only a small fraction of the total time taken to determine the modes or modal derivatives. In contrast, the factorized site matrices are stored for the duration of the algorithm, as they are much smaller and not fully dense. However, they still represent a non-negligible memory cost. To tackle very large scale problems, a more refined algorithm would perform cluster operations in a topological order to ensure that all clusters involving a specific set of sites are computed consecutively, and delete the factorized matrices once no longer necessary.

\section{Computation of Wannier derivatives}\label{sec:WannierDerivativeComputation}
In the presence of nonlinearity, the tangent stiffness matrix becomes deformation-dependent and so does the low-energy spectral subspace. Since the Wannier functions are a basis of the tangent low-energy subspace, these are also deformation dependent. To first order, we can assume that the dependence on the deformation of both the tangent stiffness matrix and and low-energy subspace is linear, with the tangent stiffness matrix having the form $K_T(q) \approx [K_{\mu \nu} + d_{i}K_{\mu \nu}]_{q=0} = K_{\mu \nu}+(\nabla_{q_i}K_{\mu \nu})dq_i$ and the local basis functions around a deformed configuration $\Psi(q_i)=\Psi_{\mu i}+\Psi'_{\mu ij}dq_j$.  To determine the Wannier derivatives $\Psi'_{\mu i j}$, we start from Eq.~\ref{eqn:WannierCoefficients}, then compute the modal sensitivities to a change $dK_{\mu \nu}$ induced by a deformation, and use the chain rule to determine the Wannier function sensitivities from the modal sensitivities. 

The sensitivity of the eigenbasis $\Psi_{\mu i}$ to changes in the tangent stiffness matrix can be calculated by determining the change in eigenvector $d\Phi_{\nu i}$ such as the eigenvalue condition, $(K_{\mu \nu}-\lambda_i M_{\mu\nu})\Phi_{\nu i}=0$, remains valid when the stiffness matrix is perturbed by $dK_{\mu \nu}$. This perturbation also have the effect of perturbing the eigenvalue by $d_j\lambda_i=\Phi_{\mu (i)}d_{j}K_{\mu \nu}\Phi_{\nu (i)}$. Expanding the eigenvalue condition yields the known expression for the modal sensitivity

\begin{equation}
(K_{\mu \nu}-\lambda_{(i)} M_{\mu \nu})d\Phi_{\nu (i)}=-(dK_{\mu \nu}-d\lambda_{(i)}M_{\mu \nu})\Phi_{\nu (i)}
\label{eqn:Modsense}
\end{equation}

In Eq.~\ref{eqn:Modsense}, the perturbation of the stiffness matrix $dK_{\mu \nu}$ appears always only contracted with an eigenvector. Therefore, we can define a \emph{nonlinear residual force} as $f_{\mu ij}=(\nabla_{q_j}K_{\mu \nu})\Phi_{\nu i}$. Using this definition, Eq.~\ref{eqn:Modsense} becomes 

\begin{equation}
\begin{split}
(K_{\mu \nu}-\lambda_{(i)} M_{\mu \nu})d_{j}\Phi_{\nu (i)}=\\ f_{\mu (i)j}-M_{\mu \nu}\Phi_{\nu (i)}(\Phi_{\tau (i)}&f_{\tau (i)j})
\end{split}
\label{eqn:ModsenseF}
\end{equation}
The usefulness of Eq.~\ref{eqn:ModsenseF} is that the nonlinear distortion of the basis functions can be calculated by evaluating a set of residual forces $f_{\mu ij}$, without the need to assembling or differentiating large matrices. This nonlinear residual force can be readily calculated using the built-in autodifferentiation capabilities of open-source packages such as FEniCSx~\cite{Scroggs2022, 10.1145/2566630, baratta_2023_10447666, 10.1145/3524456}, without the need for assembling large matrices or tensors -- by simply defining the displacement field $u_\mu=q_i\Phi_{\mu i}$ and computing the derivative of the residual force with respect to the $q$ parameters, $$ f_{\mu i j}=\frac{d^2 f_{\mu}}{dq_idq_j}.$$

The derivatives of the Wannier functions are obtained from the definition of the Wannier function (Eq.~\ref{WF} and Eq.~\ref{eqn:WannierCoefficients}) in terms of the eigenmodes, by using the chain rule with the modal sensitivity (Eq.~\ref{eqn:ModsenseF}). The problem of eigenvector sensitivities is known to be ill-conditioned, as a consequence of the gauge freedom of the eigenvectors. Typical solutions to this problem involve using a pseudoinverse to solve Eq.~\ref{eqn:ModsenseF}, or constraining the norm of the eigenvectors to remove the singularity. For the weakly-interacting, symmetric or highly-degenerate problems that typically emerge in perturbative metamaterials, the resulting system can still be singular or very poorly conditioned. This is because the left hand side of Eq.~\ref{eqn:ModsenseF}, the term $(K_{\mu \nu}-\lambda_{(i)} M_{\mu \nu})$, contains many small eigenvalues when the dynamical system has eigenvalues close to $\lambda_{(i)}$. However, these singularities only appear in the intermediate eigenvalue sensitivity calculations, not the final Wannier derivatives. This can be seen by combining Eq. ~\ref{WF} and Eq.~\ref{eqn:WannierCoefficients}):
\begin{subequations}
\begin{align}
\hat{\Psi}_{\nu (i)}&=\Phi_{\nu j}\Phi_{\mu j}M_{\mu \sigma}\xi_{\sigma i} \\
\Psi_{\mu i}&=\frac{\hat{\Psi}_{\mu(i)}}{\sqrt{\hat{\Psi}_{\nu (i)}M_{\nu \sigma} \hat{\Psi}_{\sigma (i)}}}
\end{align}
\label{eqn:WannierProjection}
\end{subequations}

Equations ~\ref{eqn:WannierProjection}a,b illustrate how the Wannier function can be understood as the mass-normalized projection of the shape function $\xi_i$ on the subspace spanned by the low-energy eigenvectors. This interpretation means that the Wannier functions are only affected by distortions that change the subspace spanned by the eigenvectors $\Phi_{\mu i}$, but not by mixing between basis elements. Since only changes $d_{j}\Phi_{\mu i}$ that are mass-orthogonal to all vectors $d\Phi_{\mu i}$ will change the space being spanned --- changes that are not orthogonal to all vectors $\Phi_{\mu i}$ can be understood as a unitary transformation of the basis vectors, that does not change the space being spanned. The computational relevance of this result is that we only need to calculate the part of $d_{j}\Phi_{\mu i}$ that is orthogonal to all eigenvectors. This allows us to constain all the small eigenvalues and hence drastically improve the conditioning of the system.

\subsection{Constraint-based pseudoinverse for the computation of the Wannier derivatives}

The goal of this section is to numerically compute the part of the solution of Eq.~\ref{eqn:ModsenseF} that is mass-orthogonal to all eigenvalues $\Phi_{\nu i}$. We do so by defining
\begin{subequations}
\begin{align}
&D=K_{\mu \nu}-\lambda_{(i)} M_{\mu \nu} \\
&x=d_{j}\Phi_{\nu (i)} \\
&y=M_{\mu \nu}\Phi_{\nu (i)}(\Phi_{\tau (i)}f_{\tau (i)j})- f_{\mu (i)j}.
\end{align}
\end{subequations}
With these definitions, Eq.~\ref{eqn:ModsenseF} becomes 
\begin{equation}
D_{\mu \nu}x_{\nu}=y_{\mu},
\end{equation}
We define the part of $x_{\nu}$ that is orthogonal to the set of eigenvectors $\Phi_{\mu i}$ of $D_{\mu \nu}$ as
\begin{equation}
\hat{x}_{\mu}=(I_{\mu \nu}-\Phi_{\mu i} \Phi_{\tau i} M_{\tau \nu})x_{\nu}
\label{eqn:orthomodes}
\end{equation}
Since $\hat{x}$ is not affected by adding combinations of $\Phi_{\mu i}$, this gives us an additional control knob to improve the conditioning of the system.
\begin{equation}
D_{\mu \nu}(\hat{x}_{\nu}+\alpha_i\Phi_{\nu i})=y_{\mu} + \beta_iM_{\mu \nu}\Phi_{\nu i}
\end{equation}

To find the modal sensitivities, we will follow a three-step process: First, we will construct and solve a constrained system that does not suffer from the conditioning issues of Eq.~\ref{eqn:ModsenseF}, i.e.,  without the singularity due to gauge freedom, as well as the small eigenvalues due to near-degenerate states in the low-energy subspace. Then, we will adjust the parameters $\alpha_i$ and $\beta_i$ to zero the reaction force (Lagrange multiplier) of the constraints. Under these conditions, the solution of the constrained system equals the solution of the unconstrained system. Finally, we will orthogonalize the result to the eigenmodes of the subspace of interest, using Eq.~\ref{eqn:orthomodes}.

\subsubsection{Construction of the constrained system}
We start by defining an eliminated system $D^e_{\mu \nu}$ where some degrees of freedom are constrained to zero. To identify degrees of freedom that are suitable to constrain the modes $\Phi_{\nu i}$, we follow the procedure described by van der Valk~\cite{vdWThesis}. The procedure starts from the first mode $\Psi_{\nu 0}$ and identifies the degree of freedom $\mu$ with the maximal displacement (in absolute value). Once this point is identified, a constraint equation, $x_{\mu}=0$ generated. Then, the modes are projected into the nullspace of the constraint, identifying combinations of motion that are still possible under the constrained system. The first of such combinations is identified as a new first mode, and the procedure is repeated to generate an additional constraint. This algorithm converges when no combination of modes $\Psi_{\nu i}$ is allowed under the generated constraints, and produces as many constraint equations as modes in the original basis.

\subsubsection{Solution of the constrained system}
Constraining the system may be understood as applying a force $\lambda_{\mu_{{}_c}}$ (Lagrange multiplier) term in the right-hand side, that acts on the constrained degrees of freedom and ensures that they have the prescribed value. By defining subsets of DOFs $\Box_{c}$ for the constrained DOFs (determined in the previous section) and $\Box_{f}$ for all other (free) DOFs, the constrained system can be expressed as:
\begin{equation}
\begin{aligned}
\begin{pmatrix}
D_{\mu_{{}_f}\nu_{{}_f}} & D_{\mu_{{}_f}\nu_{{}_c}} \\
D_{\mu_{{}_c}\nu_{{}_f}} & D_{\mu_{{}_c}\nu_{{}_c}}
\end{pmatrix}
\begin{pmatrix} 
\hat{x}_{\nu_{{}_f}} + \alpha_i\Phi_{\nu_{{}_f} i} \\
0
\end{pmatrix}= \\
\begin{pmatrix} 
y_{\mu_{{}_f}} + \beta_iM_{\mu_{{}_f}\nu}\Phi_{\nu i} \\
y_{\mu_{{}_c}} + \beta_iM_{\mu_{{}_c}\nu}\Phi_{\nu i}  + \lambda_{\mu_{{}_c}}
\end{pmatrix}
\end{aligned}
\label{eqn:fullmatrix}
\end{equation}

Equation~\ref{eqn:fullmatrix} can be solved by realizing that

\begin{equation}
\hat{x}_{\nu_{{}_f}} + \alpha_i\Phi_{\nu_{{}_f} i} =D_{\nu_{{}_f} \mu_{{}_f}}^{-1}\left(y_{\mu_{{}_f}}+\beta_i M_{\mu_{{}_{f}}\sigma}\Phi_{\sigma i}\right).
\label{eqn:getmoddev}
\end{equation} Here, the submatrix $D_{\mu_{{}_f} \nu_{{}_f}}$ does not suffer from the same conditioning problems of the full matrix $D_{\mu \nu}$ and hence the corresponding linear system can be solved using standard algorithms. To match the solution of the constrained problem to the solution of the unconstrained problem, we use the freedom in choosing $\beta_i$, to make the constrained system equivalent to the unconstrained system. This is accomplished by requiring the constraint reaction force $\lambda_{\mu_{{}_c}}$ to be zero. The reaction force condition can be determined from~\ref{eqn:fullmatrix}, resulting in the expression $D_{\mu_{{}_c}\nu_{{}_f}}\left(\hat{x}_{\nu_{{}_f}} + \alpha_i\Phi_{\nu_{{}_f} i}\right)=y_{\mu_{{}_c}} + \beta_iM_{\mu_{{}_c}\nu}\Phi_{\nu i}  + \lambda_{\mu_{{}_c}}$. Zeroing $\lambda_{\mu_{{}_c}}$ produces the equation $D_{\mu_{{}_c}\nu_{{}_f}}D_{\nu_{{}_f} \tau_{{}_f}}^{-1}\left(y_{\tau_{{}_f}}+\beta_i M_{\tau_{{}_f}\sigma}\Phi_{\sigma i}\right)=y_{\mu_{{}_c}} + \beta_iM_{\mu_{{}_c}\nu}\Phi_{\nu i}$. By reordering this equation we arrive at the following system of equations for $\beta_i$

\begin{subequations}
\begin{align}
&A_{\mu_{{}_c}i}\beta_i=b_{\mu_{{}_c}} \\
&A_{\mu_{{}_c}i}=D_{\mu_{{}_c}\nu_{{}_f}}D_{\nu_{{}_f} \tau_{{}_f}}^{-1}M_{\tau_{{}_f}\sigma}\Phi_{\sigma i} -M_{\mu_{{}_c}\nu}\Phi_{\nu i} \\
&b_{\mu_{{}_c}}=y_{\mu_{{}_c}}-D_{\mu_{{}_c}\nu_{{}_f}}D_{\nu_{{}_f} \tau_{{}_f}}^{-1}y_{\tau_{{}_f}}
\end{align}
\label{eq:systembeta}
\end{subequations}

The system of equations Eq.~\ref{eq:systembeta} can be directly solved as its dimension equals only the number of modes in the basis. Inserting the resulting $\beta_i$ into Eq.~\ref{eqn:getmoddev} and then orthogonalizing $x=\hat{x}_{\nu_{{}_f}} + \alpha_i\Phi_{\nu_{{}_f} i}$ using Eq.~\ref{eqn:orthomodes} produces the modal derivatives.

\subsubsection{Iterative solver preconditioning}
Determining the modal derivatives requires solving a total of $n(n+m)$ linear systems, where $n$ is the number of modes in the basis, and $m$ is the number of Wannier functions that need to be differentiated with respect to. These systems contain $n$ different left hand sides as the matrix $D_{\mu \nu}$ depends on the eigenvalue $\lambda_i$. A direct solution of the system would thus require $n$ matrix factorizations, a very expensive operation (note that the dimension of $D_{\mu \nu}$ corresponds to the number of degrees of freedom in a cluster). To solve this problem, we use a preconditioned iterative solver. If the effective theory corresponds to the lowest bands of a material, the matrix $D_{\mu \nu}$ is positive definite and a conjugate gradient solver can be used. If the effective theory covers some intermediate band (i.e., if there are material resonances with a frequency lower than the bands of interest, that are not included in the effective theory), the matrix $D_{\mu \nu}$ will not be positive definite and a generalized minimum residual method shall be used.

To avoid having to construct the preconditioner from scratch, we adapt the one used in the calculation of the eigenfunctions (Appendix \ref{sec:WannierComputation}), that was the inverse of the cluster stiffness matrix $K_{\mu \nu}^{-1}$. The first roadblock towards reusing this preconditioner arises from the fact that the matrix $D_{\mu_{{}_f} \nu_{{}_f}}$ appearing in Eqs.~\ref{eq:systembeta} and ~\ref{eqn:getmoddev} does not have the same dimension as $K_{\mu \nu}$. To address this problem, we construct a matrix of the form 
\begin{equation}
D^{c}_{\mu \nu}=
\begin{pmatrix}
D_{\mu_{{}_f}\nu_{{}_f}} & 0 \\
0 & \gamma I_{n}
\end{pmatrix},
\label{eqn:constrainedmatrix}
\end{equation} where $\gamma$ is a parameter chosen for numerical stability, in our case chosen to be comparable to the mean value of the diagonal of $D_{\mu \nu}$. The solution $x_{\nu_{{}_f}}=D^{-1}_{\mu_{{}_f} \nu_{{}_f}}y_{{}_f}$ can then be obtained by solving the system

\begin{equation}
\begin{aligned}
D_{\mu \nu}^c
\begin{pmatrix} 
\hat{x}_{\nu_{{}_f}}\\
\hat{x}_{\nu_{{}_c}}
\end{pmatrix} = 
\begin{pmatrix} 
\hat{y}_{\nu_{{}_f}}\\
0
\end{pmatrix}
\end{aligned}
\label{eqn:sysexpanded}
\end{equation}

Although the system in Eq.~\ref{eqn:sysexpanded} can be iteratively solved with $K_{\mu \nu}^{-1}$ as a preconditioner, we observe that the convergence of the solver is very slow, because the matrix $D_{\mu \nu}^c$ is too different from the matrix $D_{\mu \nu}$ for which the preconditioner was originally constructed. A much more suitable preconditioner is the constrained cluster stiffness matrix $(K^c)^{-1}_{\mu \nu}$, obtained by replacing $D$ by $K$ in Eq.~\ref{eqn:constrainedmatrix}. This preconditioner can be efficiently constructed by realizing that $K_{\mu \nu}^c$ only differs from $K_{\mu \nu}$ by a low-rank update, 
\begin{equation}
K_{\mu \nu}^c=K_{\mu \nu}-U_{\mu i}V^T_{i \nu},
\label{eqn:lora_expr}
\end{equation} 
with the matrices U and V defined as 
\begin{equation}
U = \begin{pmatrix}
K_{\mu_{{}_f}\nu_{{}_c}} & 0 \\
0 & I_{n}
\end{pmatrix} \,\,
V = \begin{pmatrix}
0 & K_{\mu_{{}_f}\nu_{{}_c}} \\
I_{n} & K_{\mu_{{}_c}\nu_{{}_c}}-\gamma I_{n}
\end{pmatrix},
\label{eq:loramatrices}
\end{equation}
where $I_n$ is the identity matrix of size equal to the number of constrained degrees of freedom.  The dimension of the resulting matrices $U$ and $V$ is $n\cdot 2m$ where n is the number of cluster DOFs, and $m$ is the number of constraints. After expressing the matrix $K^c_{\mu \nu}$ as a low-rank update, the updated inverse can be obtained through the Woodbury formula~\cite{guttman1946enlargement},

\begin{equation}
(K^c)^{-1}=\left( I+W\left( I-VW\right)V \right)K^{-1},
\label{eq:woodbury}
\end{equation}
where the matrix $W=K^{-1}U$. It should be noted that, although Eq.~\ref{eq:woodbury} is a low-rank update, the intermediate terms in the inverse update formula can be dense. Thus, in the solver, the matrix $(K^c)^{-1}$ (Eq.~\ref{eq:woodbury}) is implemented as a linear operator acting on a vector.

\subsubsection{Interpretation of the Wannier derivatives}
Although the introduction of the nonlinear coordinate transformation (Eq. \ref{eq:NonlinearCoordTransformation}) and Wannier derivatives has been motivated by the distortion of the basis functions as a consequence of nonlinearity, Eq.~\ref{eqn:ModsenseF} provides an alternative interpretation for the nonlinear coordinate transformation. The equation can be rewritten as
\begin{equation}
\begin{split}
\left(K_{\mu \nu}-\lambda_{(i)} M_{\mu \nu}\right)d_{j}\Phi_{\nu (i)}=\\\left( I_{\mu \tau}- M_{\mu \nu}\Phi_{\nu (i)}\Phi_{\tau (i)} \right) &f_{\tau (i)j}.
\end{split}
\label{eqn:ModsenseFRev}
\end{equation}

The right hand side in Eq.~\ref{eqn:ModsenseFRev} is the response of the system at the frequency of oscillation of the specific mode $\lambda_{(i)}$, while the right-hand side of the equation encodes the change in force due to nonlinear effects, orthogonalized to the cluster eigenmodes. Thus, the modal derivative $d_{j}\Phi_{\nu(i)}$ can be understood as encoding the displacement response to the nonlinear forces $f_{\tau(i)j}$, from all the modes that are not included in the cluster basis. 

Equation~\ref{eqn:ModsenseFRev} also highlights a key limitation of the coordinate transformation (Eq.~\ref{eq:NonlinearCoordTransformation}). The nonlinear correction term $\frac{1}{2}q_i(t)q_j(t)\Psi_{ij}'(x)$ appearing in the coordinate transformation has been calculated assuming that the excluded eigenmodes are driven at the frequencies of oscillation of $q_i$ and $q_j$, $\omega_i$ and $\omega_j$ respectively. However, the term $q_iq_j$ will contain oscillations at $\omega_i+\omega_j$ and $\omega_i-\omega_j$. The difference between the two responses will be small if the modes that are excited by the nonlinear interaction are spectrally well separated from the frequency $\omega_i+\omega_j$---as it is the case when the Wannier function encodes a slow flexural deformation while the Wannier derivative consists primarily of very fast longitudinal waves. This assumption is analogous to the Born-Oppenheimer approximation~\cite{bornop} where the response of electronic degrees of freedom is assumed to be instantaneous in comparison to nuclear degrees of freedom. However, the approximation will break down if the spectral separation condition is violated. This limitation can be overcome by introducing a velocity-dependent coordinate transformation~\cite{Gobat2023} that allows to separate the two frequency components in the oscillation of $q_iq_j$.

\section{Adjoint differentiation of the discrete model}

\begin{figure}[t]
	\centering
    \includegraphics[width=\columnwidth]{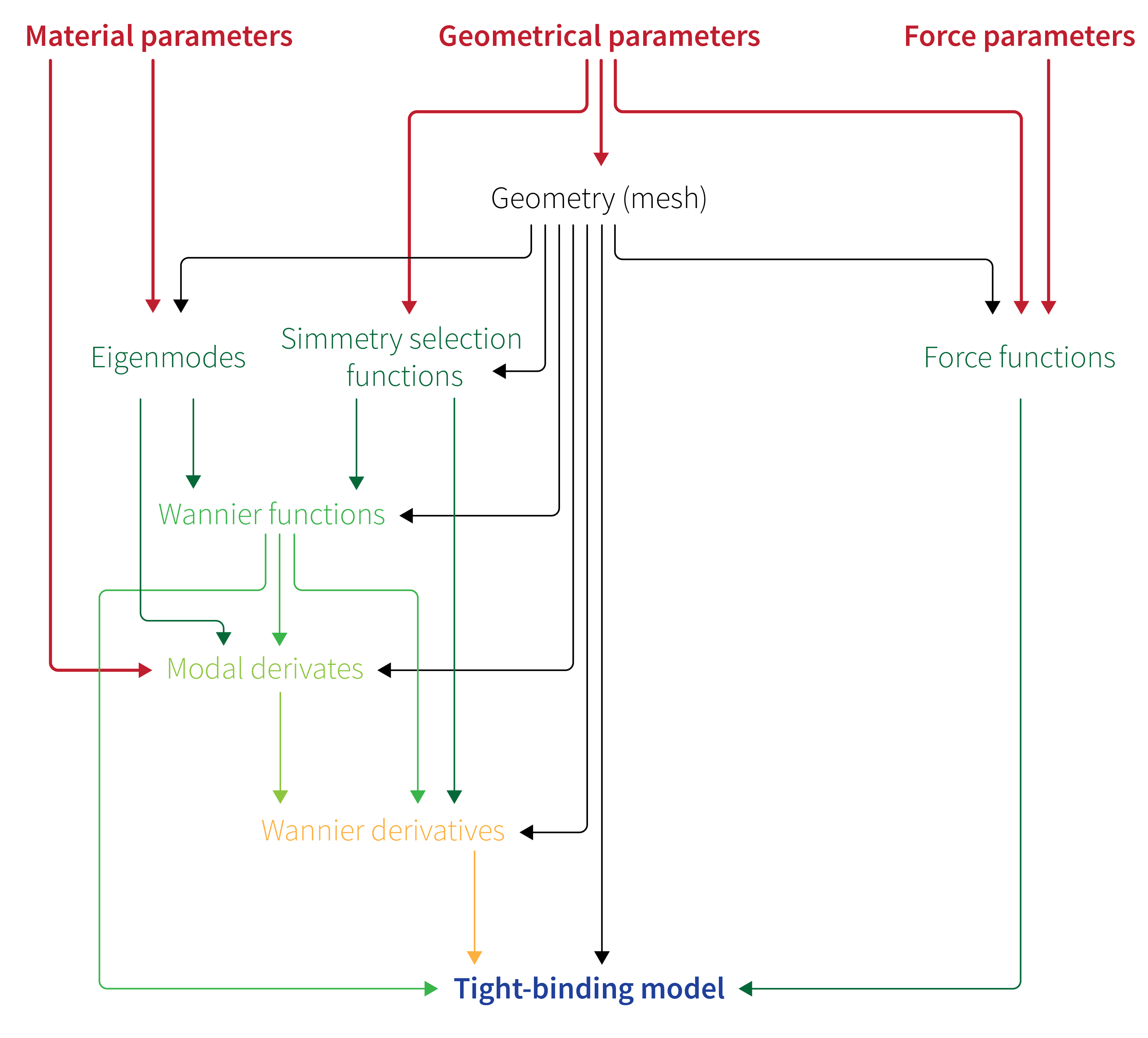}
\caption{ Reduced model computation graph | 
{\fontfamily{phv}\selectfont}
The reduced model is determined from the geometric, material and excitation force parameters following this sequence of computations. First the geometry (mesh) is determined from the geometric parameters. Then, the eigenmodes are calculated, from which the basis functions are determined using a set of projection functions. Then, the sensitivity of the eigenmodes to deformations following each basis function is calculated. These modal derivatives are then used to determine the basis function derivatives. In parallel, the force parameters are used to evaluate the force functions at every point of the mesh. Finally, the basis functions, basis function derivatives and excitation forces are used to determine the effective model. Any function that is spatially-evaluated (e.g. the force functions) or that involves a volumetric integral depends implicitly on the geometry (mesh). }
\label{fig:computegraph}
\end{figure}
A key ingredient for design and optimization is the ability to compute gradients of relevant quantities with respect to geometric parameters. The relevant quantity is typically a misfit $\eta(f, K, \Lambda, \Gamma, B, ...)$; a function of the reduced model parameters that determines the \emph{quality} of the model. This misfit may be a function of the difference between the current and desired reduced models -- if the desired target model is known in advance, or a quantity obtained by differentiating a numerical simulation of the reduced model --- for example, a speech classification accuracy obtained by simulating the system over a large dataset of spoken utterances.  

When the high number of parameters and the non-trivial cost of computing the the reduced model precludes techniques such as numerical differentiation (central differences), adjoint or reverse-mode differentiation can be employed. While in regular (forward-mode) differentiation the sensitivity of every reduced model parameter with respect to every geometric parameter is determined, in adjoint differentiation only the sensitivity of a scalar misfit function parameters is calculated (with respect to all geometric parameters). This is done by \emph{back-propagating} the sensitivity along the computation graph (Fig.~\ref{fig:computegraph}). Reverse-mode differentiation traverses the graph from the end: Starting from the sensitivity of the misfit to the model parameters, the sensitivity with respect to the basis functions, basis derivatives is calculated following the chain rule, until the sensitivity with respect to the geometric, material and force parameters is determined. When a scalar misfit is differentiated, the back-propagated sensitivities have the same dimension as the original input arguments of the function. For example, if a ROM parameter depends on $n$ Wannier functions $\Psi_{\mu i}$, backpropagating the misfit gradient will produce $n$ modal sensitivities $\nabla_{\Phi_{{}_i}}\eta$. Thus, determining the sensitivity of the misfit with respect to the geometric parameters is not significantly more computationally expensive than computing the reduced model parameters.

Finite element packages such as FEniCSx can directly differentiate integral forms, such as those appearing in the computation of the model parameters, with respect to the coordinates of the finite element mesh nodes~\cite{Ham2019}.  Since the Wannier functions do not depend on the gauge of the eigenfunctions, the derivatives of the cluster eigenfunctions can be evaluated following an approach similar to Appendix~\ref{sec:WannierDerivativeComputation}. The computation of the Wannier derivatives does not present gauge ambiguity and thus can be differentiated using the standard chain rule.


\end{document}